\documentclass[%
 aip,
jap,
 amsmath,amssymb,
 reprint,%
]{revtex4-1}

\usepackage{graphicx}
\usepackage{dcolumn}
\usepackage{bm}

\usepackage{mathrsfs}
\usepackage[utf8]{inputenc}
\usepackage[english]{babel}
\usepackage{amsmath}
\usepackage{amsfonts}
\usepackage{amssymb}
\usepackage{braket}
\usepackage{subfig}
\usepackage{soul}
\usepackage[colorlinks=true,linkcolor=blue,urlcolor=blue,citecolor=blue,breaklinks=true]{hyperref}
\usepackage[outercaption]{sidecap}
\usepackage[usenames, dvipsnames]{color}

\newcommand*{\citen}[1]{%
	\begingroup
	\romannumeral-`\x 
	\setcitestyle{numbers}%
	\cite{#1}%
	\endgroup   
}

\usepackage{soul}
\usepackage{subfig}

\begin{document}

\preprint{AIP/123-QED}

\title[]{Temperature-dependent Optoelectronic Properties of \\Quasi-2D Colloidal Cadmium Selenide Nanoplatelets}
\thanks{Ancillary document available: TE/TM mode TMEs, NPLs dimension analysis, Quasi-Fermi energy levels, Fermi factor, Intraband state gaps, Results comparison and validation, TRPL fit \& response, and comparative (experiment and theory) data tables.}

\author{Sumanta Bose}
\email{sumanta001@e.ntu.edu.sg}
\affiliation{School of Electrical and Electronic Engineering, Nanyang Technological University, 50 Nanyang Avenue, Singapore 639798, Singapore}
\affiliation{OPTIMUS, Centre for OptoElectronics and Biophotonics, Nanyang Technological University, 50 Nanyang Avenue, Singapore 639798, Singapore}

\author{Sushant Shendre}
\affiliation{School of Electrical and Electronic Engineering, Nanyang Technological University, 50 Nanyang Avenue, Singapore 639798, Singapore}
\affiliation{LUMINOUS! Centre of Excellence for Semiconductor Lighting \& Displays and TPI -- The Photonics Institute, Nanyang Technological University, Singapore 639798}

\author{Zhigang Song}
\affiliation{School of Electrical and Electronic Engineering, Nanyang Technological University, 50 Nanyang Avenue, Singapore 639798, Singapore}
\affiliation{State Key Laboratory of Superlattices and Microstructures, Institute of Semiconductors, Chinese Academy of Sciences, Beijing 100083, People's Republic of China}

\author{Vijay Kumar Sharma}
\affiliation{School of Electrical and Electronic Engineering, Nanyang Technological University, 50 Nanyang Avenue, Singapore 639798, Singapore}
\affiliation{LUMINOUS! Centre of Excellence for Semiconductor Lighting \& Displays and TPI -- The Photonics Institute, Nanyang Technological University, Singapore 639798}
\affiliation{School of Physical and Mathematical Sciences, Nanyang Technological University, 50 Nanyang Avenue, Singapore 639798, Singapore}
\affiliation{Department of Physics, Department of Electrical and Electronics Engineering and UNAM, Institute of Materials Science and Nanotechnology, Bilkent University, Turkey}

\author{Dao Hua Zhang}
\email{edhzhang@ntu.edu.sg}
\affiliation{School of Electrical and Electronic Engineering, Nanyang Technological University, 50 Nanyang Avenue, Singapore 639798, Singapore}
\affiliation{OPTIMUS, Centre for OptoElectronics and Biophotonics, Nanyang Technological University, 50 Nanyang Avenue, Singapore 639798, Singapore}
\affiliation{LUMINOUS! Centre of Excellence for Semiconductor Lighting \& Displays and TPI -- The Photonics Institute, Nanyang Technological University, Singapore 639798}

\author{Cuong Dang}
\email{hcdang@ntu.edu.sg}
\affiliation{School of Electrical and Electronic Engineering, Nanyang Technological University, 50 Nanyang Avenue, Singapore 639798, Singapore}
\affiliation{LUMINOUS! Centre of Excellence for Semiconductor Lighting \& Displays and TPI -- The Photonics Institute, Nanyang Technological University, Singapore 639798}

\author{Weijun Fan}
\email{ewjfan@ntu.edu.sg}
\affiliation{School of Electrical and Electronic Engineering, Nanyang Technological University, 50 Nanyang Avenue, Singapore 639798, Singapore}
\affiliation{OPTIMUS, Centre for OptoElectronics and Biophotonics, Nanyang Technological University, 50 Nanyang Avenue, Singapore 639798, Singapore}

\author{Hilmi Volkan Demir}
\email{hvdemir@ntu.edu.sg}
\affiliation{School of Electrical and Electronic Engineering, Nanyang Technological University, 50 Nanyang Avenue, Singapore 639798, Singapore}
\affiliation{LUMINOUS! Centre of Excellence for Semiconductor Lighting \& Displays and TPI -- The Photonics Institute, Nanyang Technological University, Singapore 639798}
\affiliation{School of Physical and Mathematical Sciences, Nanyang Technological University, 50 Nanyang Avenue, Singapore 639798, Singapore}
\affiliation{Department of Physics, Department of Electrical and Electronics Engineering and UNAM, Institute of Materials Science and Nanotechnology, Bilkent University, Turkey}



\date{\today}

\begin{abstract}
Colloidal Cadmium Selenide (CdSe) nanoplatelets (NPLs) are a recently developed class of efficient luminescent nanomaterial suitable for optoelectronic device applications. A change in temperature greatly affects their electronic bandstructure and luminescence properties. It is important to understand how-and-why the characteristics of NPLs are influenced, particularly at elevated temperature, where both reversible and irreversible quenching processes come into picture. Here we present a study on the effect of elevated temperature on the characteristics of colloidal CdSe NPLs. We used an effective-mass envelope function theory based 8-band \textit{k}$\cdot$\textit{p} model and density-matrix theory considering exciton-phonon interaction. We observed the photoluminescence (PL) spectra at various temperatures for their photon emission energy, PL linewidth and intensity by considering the exciton-phonon interaction with both acoustic and optical phonons using Bose-Einstein statistical factors. With rise in temperature we observed a fall in the transition energy (emission redshift), matrix element, Fermi factor and quasi Fermi separation, with reduction in intraband state gaps and increased interband coupling. Also, there was a fall in the PL intensity, along with spectral broadening due to an intraband scattering effect. The predicted transition energy values and simulated PL spectra at varying temperatures exhibit appreciable consistency with experimental results.  Our findings have important implications for application of NPLs in optoelectronic devices, such as NPL lasers and LEDs, operating much above room temperature.
\end{abstract}

\maketitle



\section{Introduction}

Semiconductor nanoplatelets (NPLs) are a class of atomically flat quasi two-dimensional (2D) quantum confined nanocrystals, often synthesized using wide bandgap II-VI materials. \cite{ithurria08,guzelturk14,chen14,ithurria11} Advancements in colloidal chemistry have led to the synthesis of high quality single crystal NPL samples.\cite{ithurria08,guzelturk14} It is assisted by the saturation of dangling bonds on the surface, as the organic ligands block further growth.\cite{benchamekh14} This reduces nonradiative recombination paths enhancing their optical properties. They have attracted increasing interest as they can act as cost-effective and efficient luminophores in display devices, LEDs and lasers.\cite{chen14} They are excellent candidates for such applications owing to their morphological bandgap tunability, fast fluorescence lifetime and unique optical characteristics supporting bright and tunable spectral emission with narrow full-width-at-half-maxima (FWHM) spanning the entire visible to near IR spectral range.\cite{chen14,ithurria11} They exhibit strong 1D confinement as their thickness is very small (typically few monolayers (MLs)) compared to the Bohr radius.\cite{tessier13} Also, they have smaller fluorescent lifetimes than the typical colloidal quantum dots (QDs) as a result of fast band-edge exciton recombination. NPLs have been routinely synthesized with a high quantum efficiency of about 50\%.\cite{guzelturk15} Compared to QDs, they typically have narrower emission spectra, reduced inhomogeneous broadening and suppressed Auger recombination. Recently for the first time, continuous wave laser operation has been demonstrated using colloidal NPLs.\cite{grim14}

With the recent developments and potential applications of NPLs in commercial optoelectronic devices,\cite{guzelturk15,grim14} it is imperative to study their characteristics and performances at elevated temperature -- when both reversible and irreversible luminescence quenching processes come into play.\cite{zhao12b} In this work, we study in tandem using theoretical modeling and experimental measurements, the underlying physical phenomena determining the optoelectronic characteristics of CdSe NPLs across varying temperature, above the room temperature (RT). Achtstein \textit{et al.}\cite{achtstein12} studied NPL characteristics at the cryogenic range, while there are several works studying QDs and other nanocrystals across varying temperature ranges.\cite{ji15,jing09,van15} But, a work addressing the study of NPL optoelectronic properties at elevated temperatures above RT has not been reported thus far. We begin by laying down the theoretical framework, followed by a comprehensive discussion on the obtained results to understand the physics of the NPLs at elevated temperatures, and experimental methods at the end.

\section{\label{sec:theory}Theoretical Framework}
\subsection{\label{subsec:elec-bndstr}Electronic Bandstructure}
Here we study CdSe NPLs in the zincblende (ZB) phase as they are colloidally synthesized. The surrounding dielectric medium formed by the ligands in colloidal solutions helps in the optical property enhancement by surface electronic state passivation and also to stabilize them. In our model we construct each NPL atom-by-atom in a 3D structure and use an effective mass envelope function theory approach based on the 8-band \textbf{\textit{k$\cdot$p}} method. It is used to solve the eigenvalue equation to obtain the eigenenergy values and electronic structure around the $\Gamma$-point of the Brillouin zone. We simultaneously take into account the nonparabolicity of the coupled valence band (VB) and conduction band (CB) including the split-off bands. Using the Bloch function basis of $\vert$s$\rangle\uparrow$, $\vert11\rangle\uparrow$, $\vert10\rangle\uparrow$, $\vert1-1\rangle\uparrow$, $\vert$s$\rangle\downarrow$, $\vert11\rangle\downarrow$, $\vert10\rangle\downarrow$, $\vert1-1\rangle\downarrow$, the 8-band Hamiltonian can be expressed by Eq.\ \ref{eq:hamil8x8}, where all the quadratic, linear and independent terms of \textit{k} are in the first Hamiltonian matrix, with $H_{ij}=H_{ji}^\ast$, denoted by \textit{c.c.} standing for complex conjugate. \cite{chen09} $E_p$ is the Kane's theory matrix element and $V_{\text{NPL}}$ is the NPL confining potential. Expressions for Hamiltonian elements of Eq. \ref{eq:hamil8x8}, A through E are detailed below. The modified wavevectors, $k_x^\prime$, $k_y^\prime$, $k_z^\prime$ are calculated using the $3\times3$ strain tensor matrix, $\varepsilon$.

\begin{widetext}
\begin{eqnarray} \label{eq:hamil8x8}
H_{8\times8}=\left( \begin{array}{cccccccc}
		A & \frac{i\hbar\sqrt{E_p}\left(k_x^\prime+ik_y^\prime\right)}{2\sqrt{m_0}} & i\hbar\sqrt{\frac{E_p}{2m_0}}k_z^\prime & \frac{i\hbar\sqrt{E_p}\left(k_x^\prime-ik_y^\prime\right)}{2\sqrt{m_0}} & 0 & 0 & 0 & 0 \\
		c.c. & B & C & D & 0 & 0 & 0 & 0 \\
		c.c. & c.c. & E & C & 0 & 0 & 0 & 0 \\
		c.c. & c.c. & c.c. & B & 0 & 0 & 0 & 0 \\
		c.c. & c.c. & c.c. & c.c. & A & \frac{i\hbar\sqrt{E_p}\left(k_x^\prime+ik_y^\prime\right)}{2\sqrt{m_0}} & i\hbar\sqrt{\frac{E_p}{2m_0}}k_z^\prime & \frac{i\hbar\sqrt{E_p}\left(k_x^\prime-ik_y^\prime\right)}{2\sqrt{m_0}} \\
		c.c. & c.c. & c.c. & c.c. & c.c. & B & C & D \\
		c.c. & c.c. & c.c. & c.c. & c.c. & c.c. & E & C \\
		c.c. & c.c. & c.c. & c.c. & c.c. & c.c. & c.c. & B \\
		  \end{array} \right) \nonumber\\+ H_{so} + V_{\text{NPL}}
\end{eqnarray}
\end{widetext}


\begin{subequations}
\label{eqn:hamil-elements}

\begin{equation}
	A=-\dfrac{\hbar^2}{2m_0}\gamma_c\left(k_x^2+k_y^2+k_z^2\right)+a_c\left[tr\left(\varepsilon\right)\right]+E_g
\end{equation}

\begin{eqnarray}
	B=-\dfrac{\hbar^2}{2m_0}\left[\dfrac{L^\prime+M^\prime}{2}\left(k_x^2+k_y^2\right)+M^\prime k_z^2\right]
    \nonumber\\ + a_v\left[tr\left(\varepsilon\right)\right] + \frac{b}{2}\left[tr\left(\varepsilon\right)-2\varepsilon_{zz}\right]\text{\hspace{10mm}}
\end{eqnarray}

\begin{equation}
	C=-\dfrac{\hbar^2}{2m_0}\left[\dfrac{N^\prime\left(k_x-ik_y\right)k_z}{\sqrt{2}}\right]+\sqrt{6}d\left(\varepsilon_{xz}-i\varepsilon_{yz}\right)
\end{equation}

\begin{eqnarray}
	D=-\dfrac{\hbar^2}{2m_0}\left[\dfrac{L^\prime-M^\prime}{2}\left(k_x^2-k_y^2\right)-iN^\prime k_xk_y\right]
	\nonumber\\ - 
	i\sqrt{12}d\varepsilon_{xy}+\frac{3b}{2}\left(\varepsilon_{xx}-\varepsilon_{yy}\right)\text{\hspace{17mm}}
\end{eqnarray}

\begin{eqnarray}
	E=-\dfrac{\hbar^2}{2m_0}\left[M^\prime\left(k_x^2+k_y^2\right)+L^\prime k_z^2\right]
	\nonumber\\ + a_v\left[tr\left(\varepsilon\right)\right] + b\left[3\varepsilon_{zz}-tr\left(\varepsilon\right)\right]\text{\hspace{0.5mm}}
\end{eqnarray}

\begin{equation}
	\gamma_c=-\dfrac{E_p}{3}\left[\dfrac{1}{E_g+\Delta_{so}}+\dfrac{2}{E_g}\right]+\dfrac{m_0}{m_e^*}
\end{equation}

\begin{equation}
	\left(
	\begin{array}{c}
	k_x^\prime \\ k_y^\prime \\ k_z^\prime
	\end{array}
	\right) = \left(I_3-\varepsilon\right)
	\left(
	\begin{array}{c}
	k_x \\ k_y \\ k_z
	\end{array}
	\right) 
\end{equation}

\end{subequations}

Using the Luttinger-Kohn effective mass parameters $\gamma_1$, $\gamma_2$, $\gamma_3$ as enlisted in Table \ref{tab:mat-param}, we have derived the modified Luttinger parameters for our calculation.

\begin{subequations}
\label{eqn:L-K-param-eqns}
\begin{equation}
	L^\prime=L-\dfrac{E_p}{E_g}=-\dfrac{\hbar^2}{2m_0}(\gamma_1+4\gamma_2+1)
\end{equation}

\begin{equation}
	M^\prime=M=-\dfrac{\hbar^2}{2m_0}(\gamma_1-2\gamma_2+1)
\end{equation}

\begin{eqnarray}
	N^\prime=N-\dfrac{E_p}{E_g}=-\dfrac{\hbar^2}{2m_0}(6\gamma_3)
\end{eqnarray}

\end{subequations}

\noindent$H_{so}$, given by Eq. \ref{eq:Hso} is the VB spin-orbit coupling Hamiltonian.

\begin{eqnarray} \label{eq:Hso}
H_{so}=\dfrac{\Delta_{so}}{3}
\left( \begin{array}{cccccccc}
		0 & 0 & 0 & 0 & 0 & 0 & 0 & 0 \\
		0 & 0 & 0 & 0 & 0 & 0 & 0 & 0 \\
		0 & 0 & -1 & 0 & 0 & -\sqrt{2} & 0 & 0 \\
		0 & 0 & 0 & -2 & 0 & 0 & \sqrt{2} & 0 \\
		0 & 0 & 0 & 0 & 0 & 0 & 0 & 0 \\
		0 & 0 & -\sqrt{2} & 0 & 0 & -2 & 0 & 0 \\
		0 & 0 & 0 & \sqrt{2} & 0 & 0 & -1 & 0 \\
		0 & 0 & 0 & 0 & 0 & 0 & 0 & 0 \\
		  \end{array} \right)
\end{eqnarray}

\noindent Considering the NPL periodicities to be $L_{x}, L_{y}, L_{z}$ along the \textit{x}, \textit{y} and \textit{z} directions, we use plane waves to expand the eight-dimensional hole and electron envelope wave function as\cite{song16}

\begin{subequations}
 \begin{equation}
 \phi _{m}=\left\{ \phi _{m}^{j}\right\} \hspace{0.5cm}(j=1, 2, ..., 8)
 \end{equation}
 with
 \begin{equation}\label{}
 \phi _{m}^{j}=\frac{1}{\sqrt{V}} \sum_{n_{x},n_{y},n_{z}}a_{m,n_{x},n_{y},n_{z}}^{j}e^{
 [i(k_{nx}x+k_{ny}y+k_{nz}z)]}
 \end{equation}
\end{subequations}

\noindent where $k_{nx}=2\pi n_{x}/L_{x}$, $k_{ny}=2\pi n_{y}/L_{y}$, $k_{nz}=2\pi n_{z}/L_{z}$ and $n_{x},n_{y},n_{z}$ are the plane wave numbers. And $V=L_{x}L_{y}L_{z}$. The indices of the basis and energy subbands are given by \textit{j} and \textit{m} respectively.

Now, the interatomic interactions in the NPL cause atomistic relaxations which can be estimated by employing the microscopic theory of a valence force field (VFF) model, wherein we account for 2-body interaction (atomic-bond-stretching) and 3-body interaction (atomic-bond-bending). The sum total of the strain energy is given by\cite{bose14b}

\begin{eqnarray}
	E_{\text{VFF}}=\sum\limits_{i(j)}^{ }\dfrac{3\alpha_{ij}}{16d_{0,ij}^2}\bigg(\vert\textbf{r}_i-\textbf{r}_j\vert^2-d_{0,ij}^2\bigg)^2+\sum\limits_{i(j,k)}^{ }\dfrac{3\beta_{jik}}{8d_{0,ij}d_{0,ik}} \nonumber\\
	\bigg(\vert\textbf{r}_i-\textbf{r}_j\vert\vert\textbf{r}_i-\textbf{r}_k\vert-\cos\hat{\theta}_{ijk}\cdot d_{0,ij}d_{0,ik}\bigg)^2 \text{\hspace{6mm}}
	\label{eqn:vff-SE-final-2}
\end{eqnarray}

\noindent where the atoms in the crystal lattice are identified by the indices \textit{i}, \textit{j} and \textit{k}. $d_{0,ij}$ is the ideal atomic bond-length between the $i^{th}$ and $j^{th}$ atoms, with $4 d_{0,ij}=\sqrt{3} a_{0,ij}$ relating it to $a_{0,ij}$, the lattice constant.\cite{bose14a} Factors such as external piezoelectric strain or temperature cause deviations in the bond-length from the ideal $d_{0,ij}$, and the bond-distance between the $i^{th}$ and $j^{th}$ atoms in the system studied is given by $\vert\textbf{r}_i-\textbf{r}_j\vert$, where $\textbf{r}_i$ is the position vector of the $i^{th}$ atom. The bond-stretching force constant of the $i-j$ bond is given by $\alpha_{ij}$ (Table \ref{tab:mat-param}). The ideal bond-angle of the bond among the $i^{th}$, $j^{th}$ and $k^{th}$ atoms (vertexed at $i^{th}$ atom) is given by $\hat{\theta}_{ijk}$ ($=\frac{1}{3}$ for ZB structures). The bond-bending force constant of the $j-i-k$ bond-angle is given by $\beta_{jik}$ (Table \ref{tab:mat-param}).

\begin{table}
\caption{\label{tab:mat-param}Physical Material Parameters of CdSe at Room Temperature}
\begin{ruledtabular}
\begin{tabular}{lcr}
Symbol & Physical parameter & Value$^{\text{ref.}}$\\
\hline
$E_g$ (eV) & Bandgap energy & 1.732$^a$ \\
    $E_p$ (eV) & Kane Matrix Element & 16.5$^a$ \\
    $\Delta_{so}$ (eV) & Spin-orbit splitting energy & 0.42$^b$ \\
    $\gamma_1$ & & 3.265$^c$ \\
    $\gamma_2$ & Luttinger-Kohn parameters  & 1.162$^c$\\
    $\gamma_3$ & & 1.443$^c$ \\
    $m_e^\ast/m_0$ & Effective Electron Mass & 0.12$^d$ \\
    $m_h^\ast/m_0$ & Effective Hole Mass & 0.45$^d$ \\
    $a_c$ (eV) & Hydrostatic deformation potential of CB & -2.83$^a$ \\
    $a_v$ (eV) & Hydrostatic deformation potential of VB & 1.15$^a$ \\
    $b$ (eV) & Sheer deformation potential  & -1.05$^a$ \\
    $d$ (eV) & Sheer deformation potential  & -3.10$^a$ \\
    $a_0$ (\AA) & Lattice constant & 6.052$^d$ \\
    $d_0$ (\AA) & Interatomic bond-length & 2.620$^e$ \\
    $\alpha$ (N/m) & Bond-stretching force constant & 6.05$^e$ \\
    $\beta$ (N/m) & Bond-bending force constant & 6.052$^e$ \\
	$n_r$ & Refractive Index & 2.5$^a$\\
    $a_B$ (nm) & Bohr radius & 5.6$^f$\\
    \hline
$^a$Ref.\ \citen{bose16}\hspace{-0.1cm}&$^b$Ref.\ \citen{shan94}\hspace{0.16cm}$^c$Ref.\ \citen{karazhanov05}\hspace{0.16cm}$^d$Ref.\ \citen{madelung}\hspace{0.16cm}$^e$Ref.\ \citen{kumar00}\hspace{-0.1cm}&$^f$Ref.\ \citen{landry14}\\
\end{tabular}
\end{ruledtabular}
\end{table}

\subsection{\label{subsec:opt-prop}Optical Properties}
For studying the optical characteristics of NPLs, we use the density-matrix equation dependent on the quasi Fermi levels of the CB ($f_c$) and VB ($f_v$). We can estimate it for a two-dimensional structure with exciton effects as a sum of the spontaneous radiative rate from the excitonic bound states, $r_{sp}^{ex,b}$ and continuum-states, $r_{sp}^c$. The excitonic bound state contribution is given by\cite{micallef93,herbert92}

\begin{subequations}\label{eq:ex}
	\begin{eqnarray}\label{rspb}
		r_{sp}^{ex,b}\left(E\right)=\dfrac{e^2n_rE}{\pi m_0^2\varepsilon_0c^3\hbar^2t}\sum\limits_{c,v}\vert\Psi_{1s}^{cv}\left(0\right)\vert^2\vert\mathcal{P}_{cv}\vert^2\cdot\vert {I}_{cv}\vert^2\times\nonumber\\
		f_c\left(1-f_v\right)\mathscr{L}(E-E_{cv}-E_b)\text{\hspace{10mm}}
	\end{eqnarray}
	\begin{equation}
		\Psi_{1s}\left(x\right)=\frac{4\beta}{a_{B}\sqrt{2\pi}}e^{-2x\beta/a_B}
	\end{equation}
	\begin{equation}
    \label{eq:ex-bind-EN}
		E_b=-4\beta^2R_y
	\end{equation}
\end{subequations}

\noindent where symbols (\textit{e}, $n_r$, $E$, $m_0$, $\varepsilon_0$, $c$, $\hbar$) have standard physical meanings. $t$ is the NPL thickness, $\Psi_{1s}\left(x\right)$ is the 1S exciton envelope function and \textit{x} is the relative distance between the electron and hole along the transverse direction in the NPL. $f_c$ and $f_v$ are the Fermi-Dirac distributions of the CB and VB, dependent on the quasi-Fermi energy levels of the CB ($E_{fc}$) and VB ($E_{fv}$), respectively. ${I_{cv}}=\int\Psi^{E(c)}\left(z\right)\Psi^{H(v)}\left(z\right)dz$ is the overlap integral between the electron and hole wavefunctions along the transverse direction. $E_b$ is the 1S exciton binding energy and $\beta$ is a variational parameter, taken to be 1 for two-dimensional structures, based on conclusive observation that for two-dimensional semiconductors, the binding energy increment by the low dielectric constant of the surface ligands and the solvent is is almost exactly compensated by self-interactions of electrons and holes with self-image potential.\cite{ithurria11} $a_B=4\pi\varepsilon_0\varepsilon_r\hbar^2/m_re^2$ and $R_y=m_re^4/32\pi^2\varepsilon_0^2\varepsilon_r^2\hbar^2$ are the excitonic Bohr radius and excitonic Rydberg energy respectively, where $m_r$ is the reduced mass of the electron-hole pair: $m_r^{-1}=m_e^{-1}+m_h^{-1}$ and $\varepsilon_r$ is the relative permittivity. For the continuum state contributions, we have\cite{herbert92,Chuang}

\begin{subequations}\label{eq:c}
	\begin{eqnarray}\label{rspc}
		r_{sp}^c\left(E\right)=\dfrac{e^2n_rE}{\pi m_0^2\varepsilon_0c^3\hbar^2V}\sum\limits_{c,v}\vert\mathcal{P}_{cv}\vert^2f_c\left(1-f_v\right)\times\nonumber\\
		S_{2D}\left(E-E_{cv}\right)\mathscr{L}(E-E_{cv})\text{\hspace{10mm}}
	\end{eqnarray}
	\begin{equation}
		S_{2D}\left(E-E_{cv}\right)=\frac{2}{1+\text{exp}\left(-2\pi\sqrt{R_y/(E-E_{cv})}\right)}
	\end{equation}
\end{subequations}

\noindent where $S_{2D}\left(E\right)$ is the 2D Sommerfeld enhancement factor.\cite{Chuang} $V$ is the NPL volume in real space. The PL emission is maximum when $f_c=1$ (fully occupied upper level) and $f_v=0$ (fully empty lower level), and the emission rate is $\propto f_c\left(1-f_v\right)$, which is called the Fermi factor.\cite{piprek-book03} The calculation of the theoretical PL also takes into consideration the spectral transition energy broadening (dephasing) effect accounted by the Lorentzian broadening lineshape term $\mathscr{L}_{cv}(E-E_{cv})$ in Eq. \ref{rspb} and \ref{rspc}, given by
\begin{equation}
	\mathscr{L}_{cv}(E-E_{cv})=\dfrac{1}{\pi}\dfrac{\hbar/\tau_{in}}{\left(E-E_{cv}\right)^2+\left(\hbar/\tau_{in}\right)^2}
\label{eq:lorentz}
\end{equation}
\noindent whose full width at half maximum (FWHM) is $2\hbar/\tau_{in}$ and scattering probability per unit time is $1/\tau_{in}$, where $\tau_{in}$ is the intraband relaxation time. The term $\vert\mathcal{P}_{cv}\vert^2$ in Eq. \ref{rspb} and \ref{rspc} is the square of the optical transition matrix element (TME). It quantifies the strength of transition between the electron- and hole-subband.\cite{sugawara-book99} We calculate it using the momentum operator, \textbf{p} and the real electron and hole wavefunctions, $\Psi_{c,\textbf{\text{k}}}$ and $\Psi_{v,\textbf{\text{k}}}$ respectively: $\mathcal{P}_{cv,i}=\bra{\Psi_{c,\textbf{\text{k}}}}\hat{\text{e}}\cdot\textbf{\text{p}}_{i}\ket{\Psi_{v,\textbf{\text{k}}}}$, $i=x,y,z$.\cite{fan96}
Expressions for $\mathcal{P}_{cv,i}$ along the \textit{x}, \textit{y} and \textit{z} directions are given in the ancillary document. The average of $\mathcal{P}_{cv,x}$ and $\mathcal{P}_{cv,y}$ (TMEs along \textit{x} and \textit{y} directions) contributes to the transverse electric (TE) mode emission, while $\mathcal{P}_{cv,z}$, the TME along the \textit{z} direction contributes to the transverse magnetic (TM) mode emission. The TE mode emission is \textit{x-y} plane polarized, while the TM mode emission is \textit{z} direction polarized. Table \ref{tab:mat-param} enlists the material parameters of CdSe used in this work, citing their sources. Now, in Sec.\ \ref{sec:res-disc}, using this theoretical framework, we study the effect of temperature on the optoelectronic properties of CdSe NPLs and compare with experimental results.

\section{\label{sec:res-disc}Results and Discussions}

\subsection{\label{subsec:theor-res}Theoretical Results}

       		\begin{figure}[t]
			\centering
 \includegraphics[width=0.5\textwidth]{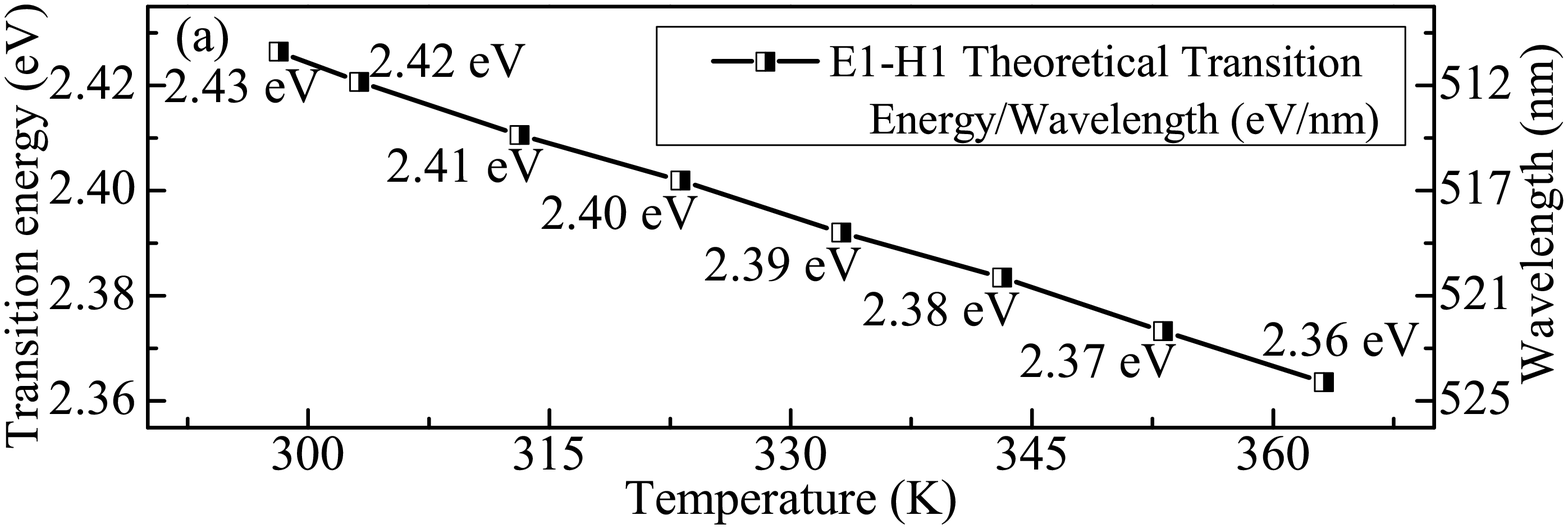}\\	\includegraphics[width=0.5\textwidth]{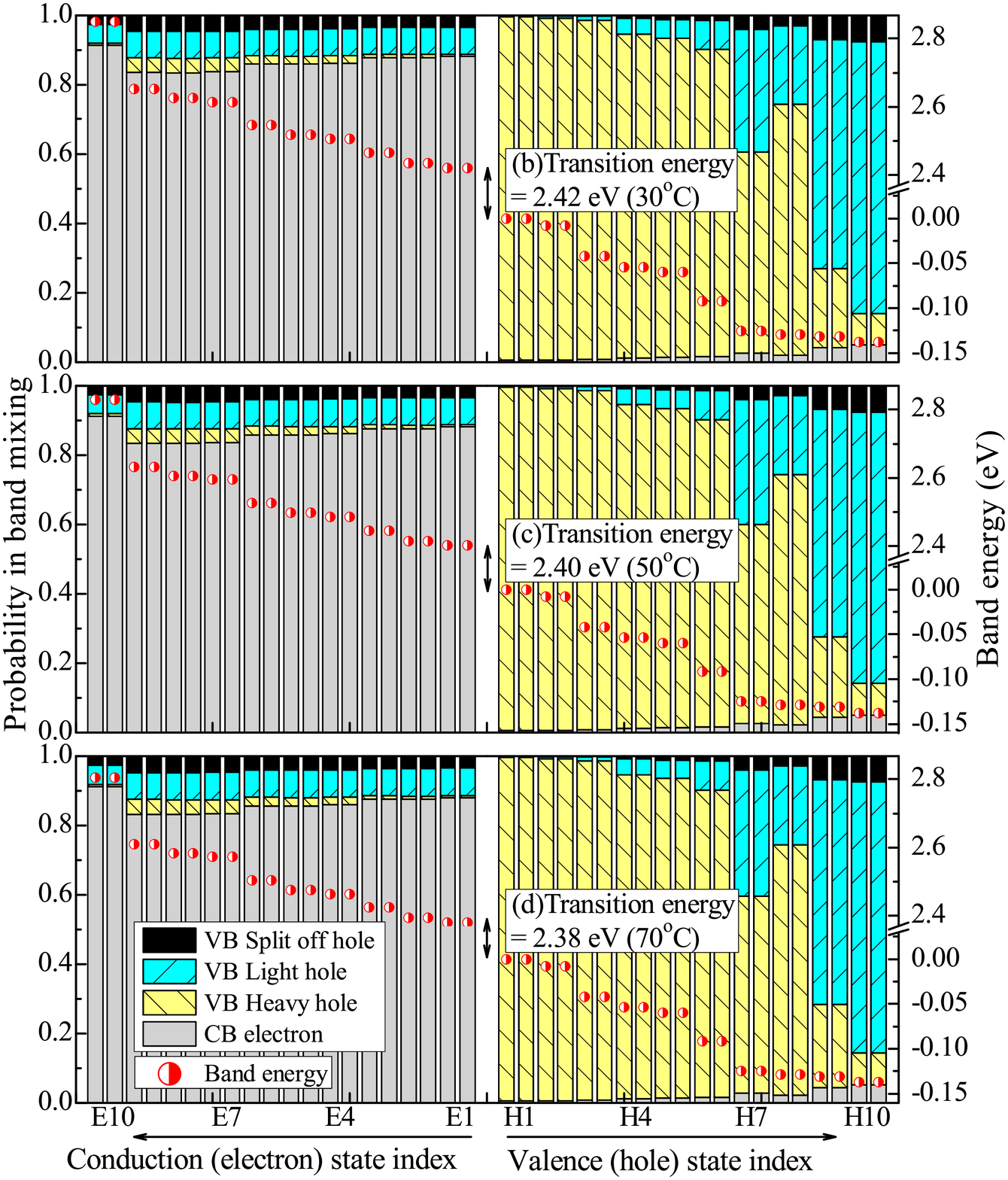}
           
			\caption{\textcolor{black}{(a: \textit{first}) Theoretical E1--H1 (bottom of conduction band to top of valence band) excitonic transition energy/wavelength as a function of temperature.} Frames (b--d) show the electronic bandstructure and the probability in band mixing between conduction electrons and valence heavy-, light- and split off holes for 4 ML CdSe NPLs at (b: \textit{second}) 30$^\circ$C, (c: \textit{third}) 50$^\circ$C and (d: \textit{fourth}) 70$^\circ$C. E1--H1 transition energies are indicated. Legends show the band mixing probabilities. }
		\label{fig:EN_prob-tempr}
		\end{figure}

We have used the theoretical model described in Sec.\ \ref{sec:theory} to study the optoelectronic characteristics of quasi two-dimensional colloidal CdSe NPLs of lateral dimensions 22 nm $\times$ 8 nm, and vertical thickness of 4 monolayer (ML), across temperature varying from room temperature (RT) to 90$^\circ$C. We have modeled and studied NPLs of this particular dimension, based on experimental measurements as we shall describe later in Sec.\ \ref{subsec:expt-res}. Under this scheme, we have studied how temperature affects their electronic structure and optical properties and predicted some temperature-dependent photoluminescence (PL) characteristics of CdSe NPLs. Fig.\ \ref{fig:EN_prob-tempr}a shows the theoretically calculated excitonic transition energy (and photon emission wavelength) from the bottom of CB (E1) to the top of VB (H1) i.e. E1-H1, as a function of temperature. With an increase in temperature there is a red shift in the transition energy (photon emission energy). The reason is that the bandgap, $E_g$ has a negative thermal coefficient (\textit{d}$E_g$\textit{/dT}) due to (\textit{i}) lattice thermal dilatation and electron-phonon interaction $\propto T$ at high temperatures, and (\textit{ii}) electron-phonon interaction $\propto T^2$ at low temperatures. In this context, we conventionally have the Varshni relation \cite{varshni67} to explicate the behavior of $E_g$. However, detailed systematic study for II-VI semiconductors have shown that sometimes the Varshni relation may not be very accurate.\cite{calderon} Another semiempirical expression suggested by Cardona \textit{et al.} \cite{logothetidis86} is more accurate as we shall subsequently see. Also, with an increase in temperature, the phonon concentration increases and causes increased scattering and non-radiative recombinations start occurring,\cite{burov07} which causes the E1-H1 TME to decrease.

       		\begin{figure}[t]
			\centering
			\includegraphics[width=0.48\textwidth]{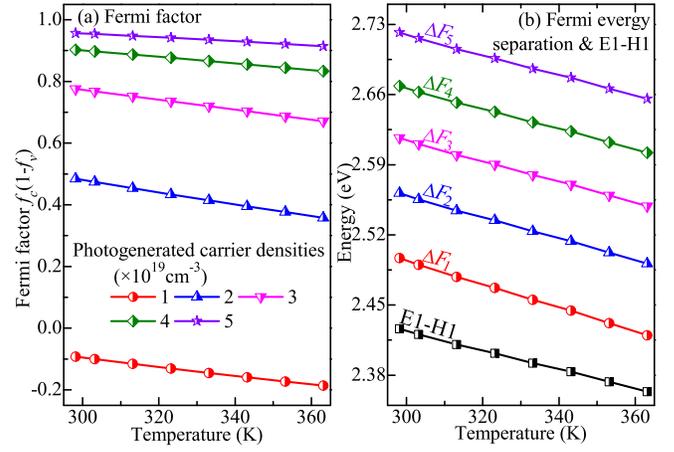}
			\caption{(a: \textit{left}) Fermi factor $f_c(1-f_v)$ for E1--H1 transition, and (b: \textit{right}) Fermi energy separation $\Delta F=E_{fc}-E_{fv}$ compared with the E1-H1 transition energy in 4 ML CdSe NPLs as a function of temperature for varying injection carrier concentration. Both, Fermi factor and $\Delta F$ have a negative temperature gradient.}
		\label{fig:FF-delF-vs-T}
		\end{figure}

     		\begin{figure}[t]
			\centering
			\includegraphics[width=0.48\textwidth]{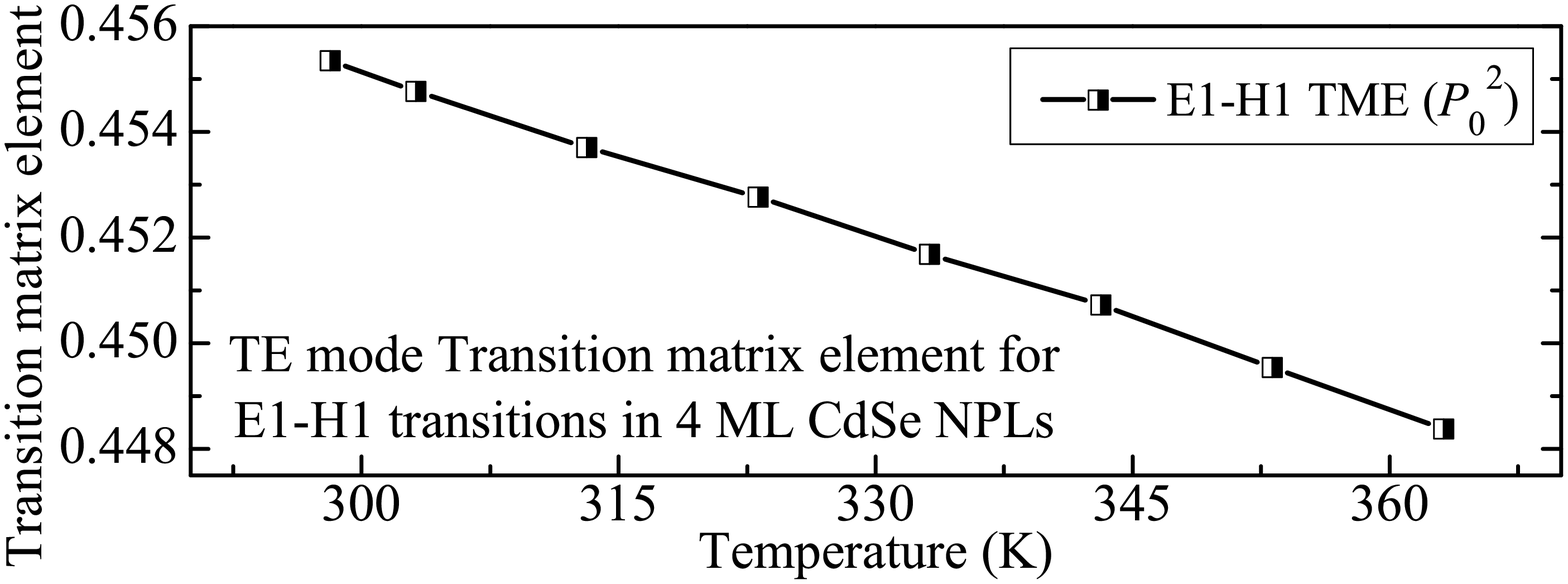}
			\caption{TE mode transition matrix element (TME) for the E1--H1 transitions as a function of temperature in 4 ML CdSe NPL. The other interband TMEs (both TE and TM mode) are given in the ancillary document.}
		\label{fig:TME}
		\end{figure}

\begin{SCfigure*}
\includegraphics[trim={4.4cm 0.4cm 1.9cm 0.3cm},width=0.43\textwidth]{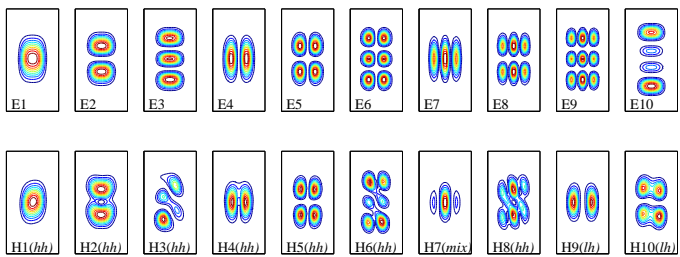}
\caption{The spatial charge density of the first ten electron and hole states of 4 ML CdSe NPLs at 30$^\circ$C (303.15 K) cut along the vertically central \textit{x-y} plane ($z=0$) of the NPL. Warmer (reddish) colors depict higher occupation probability over cooler ones (bluish). The rectangular boundary depicts our studied NPL (22 nm $\times$ 8 nm), but the dimensions are not to scale, for the sake of comprehensive visual representation.}\label{fig:wfs-cdse22_8_4_300}
\end{SCfigure*}

\begin{figure}[t]
			\centering
			\includegraphics[scale=0.3]{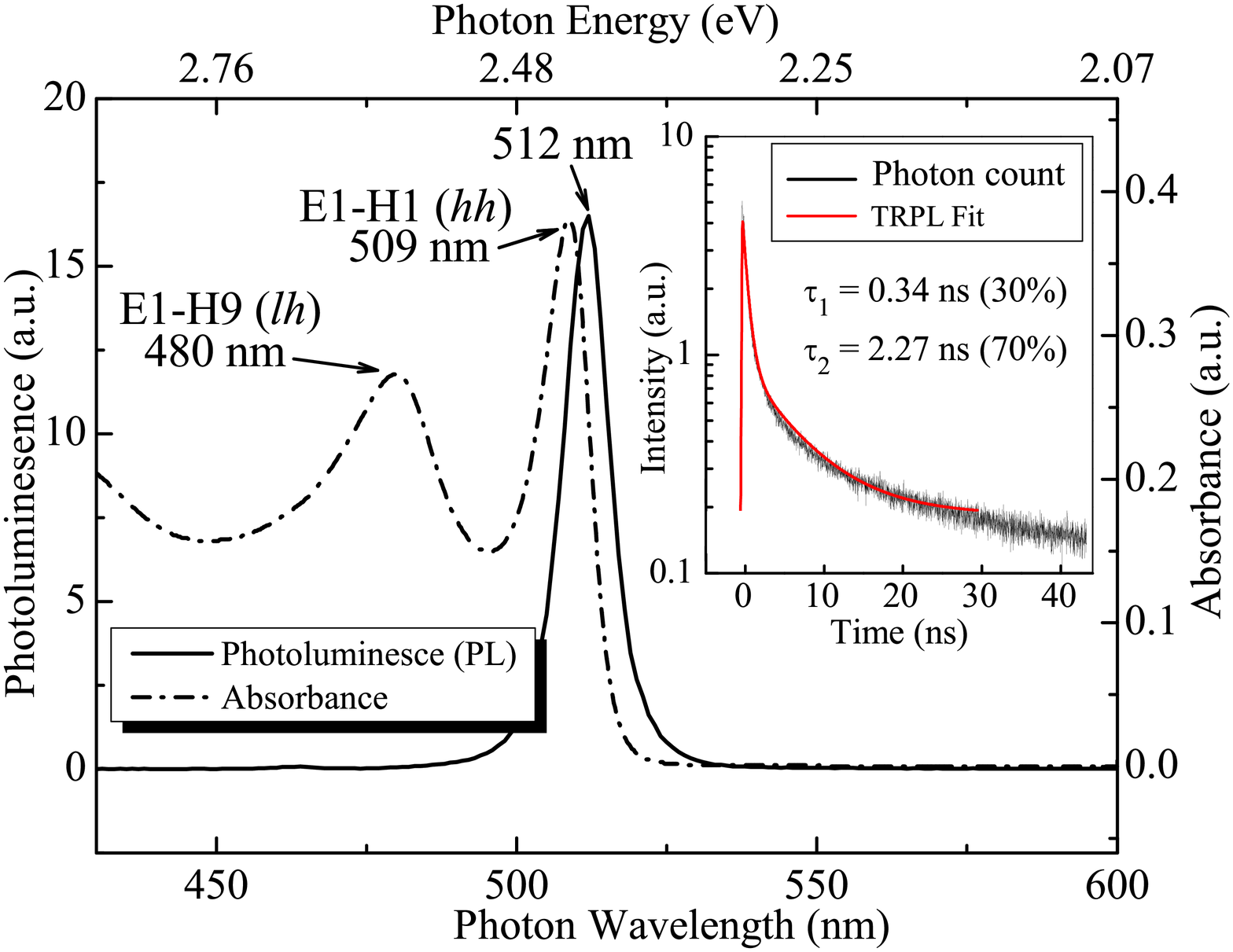}
			\caption{Experimental Photoluminescence (PL) and Absorption spectrum of 4 ML CdSe NPLs at 30$^\circ$C (303.15 K). The PL excitation wavelength was 350 nm. Both PL and absorption were measured in solution form. PL peak is at 512 nm. Primary absorption peak, at 509 nm has a Stokes shift of 3 nm, while the secondary absorption peak is at 480 nm. Inset shows Time-Resolved PL (TRPL) spectrum and fit measured in thin-film form at 30$^\circ$C. The fit and the system response is given in the ancillary document.}
		\label{fig:PL-Abs-TRPL}
		\end{figure}

Of the eight temperature cases studied in Fig.\ \ref{fig:EN_prob-tempr}a, we present the electronic bandstructure and the probability in band mixing between conduction electrons and valence heavy-, light- and split off-holes due to coupling for the 4 ML CdSe NPLs at 30, 50 and 70$^\circ$C, in Fig.\ \ref{fig:EN_prob-tempr}b-d. The E states represent the conduction electron states, while the H states represent the valence hole states. Each of the states is two fold degenerate, considering spin-up and spin-down alternatives. For the sake of discussion, let us consider the 4 ML CdSe NPL at 30$^\circ$C of Fig.\ \ref{fig:EN_prob-tempr}a as a case. The E1-H1 transition energy (2.42 eV) determines the PL peak emission energy.\footnote{Later in Sec.\ \ref{subsec:expt-res} we will show that the experimental PL emission spectrum of 4 ML CdSe NPL at 30$^\circ$C (Fig.\ \ref{fig:PL-Abs-TRPL}), has its peak emission at 512 nm corresponding to the transition energy of 2.42 eV predicted by our model.} The CB states mainly comprise of electrons with very low composition of \textit{hh}, \textit{lh} and split-off (\textit{so}) hole, which comes from the coupling effects. The effective \textit{hh} mass is much larger than the effective electron mass for CdSe, \cite{madelung} thus the \textit{hh} dispersion curve is more flat than the conduction dispersion curve. Therefore, the first ten hole levels span only $\sim$140 meV, while the first ten electron levels span $\sim$425 meV. In the VB, the first few levels are \textit{hh} dominated followed by increased influence of \textit{lh} and \textit{so}. Comparing the span of first ten electron and hole states and band-mixing probabilities of the three cases in Fig.\ \ref{fig:EN_prob-tempr} we have two significant observations: (\textit{i}) an increase in temperature induces a faster reduction in the intra-CB state gaps compared to the intra-VB state gaps. For instance, a temperature rise of 20$^\circ$C between adjacent frames of Fig.\ \ref{fig:EN_prob-tempr}, the $\Delta_{\text{E1-E10}}$ is over five times larger than the $\Delta_{\text{H1-H10}}$. This is by virtue of the temperature-induced thermal lattice dilatation effects, given than the deformation potential of the CB is higher than that of the VB for CdSe; and (\textit{ii}) an increase in temperature promotes \textit{e-h} quantum state coupling -- the conduction states start to have an increased contribution from holes and vice versa. Moreover, the intraband dynamics is also affected, such that the role of the dominant contributor diminishes with increasing temperature. For instance, the dominant contributors for H1 through H8 is \textit{hh} while for H9 and H10 is \textit{lh}, whose contribution falls as temperature rises. The contrast is however small to be reflected graphically. These band-mixing probabilities are affected by the varying degree coupling between the conduction electrons and valence \textit{hh}, \textit{lh} and \textit{so} holes, which can be empirically determined. For each level, the band-mixing probabilities are dictated by their aggregate electron and hole wavefunctions and associated charge densities thereof.

The Fermi factor and energy levels are important determiners of the temperature-dependent optoelectronics characteristics. In Fig.\ \ref{fig:FF-delF-vs-T}a we show the Fermi factor $f_c(1-f_v)$ at E1--H1 transition and in Fig.\ \ref{fig:FF-delF-vs-T}b the Fermi energy separation $\Delta F=E_{fc}-E_{fv}$ in our 4 ML CdSe NPL as a function of temperature, for varying photogenerated carrier densities from 1 to 5$\times10^{19}$ cm$^{-3}$. The $\Delta F$ for each case is compared with the E1--H1 transition energy, and $\Delta F$>E1--H1 ensures the Bernard-Duraffourg inversion condition (population inversion) necessary for lasing.\cite{Chuang} With an increase in carrier density, both Fermi factor and Fermi energy separation increases. However, as we continue to increase the density, the extent of Fermi factor increment falls, at it approaches saturation. However, a rise in temperature has detrimental effects on both, Fermi factor and Fermi energy separation. With an increase in temperature, the fermions (electrons and holes) get thermally excited and therefore the probability of occupying higher CB and VB energy states is increased; so the $f_c$ falls, while $f_v$ rises. Consequently the Fermi factor decreases with temperature. Also, the quasi Fermi levels of the CB and VB ($E_{fc}$ and $E_{fv}$) approach the band edges, and $\Delta F$ falls. In the ancillary document, we have shown the $E_{fc}$ and $E_{fv}$ in contour forms along with the Fermi factor and Fermi energy separation.

Another critical temperature-dependent optoelectronic factor is the transition matrix element (TME) which is shown in Fig.\ \ref{fig:TME} for the E1--H1 transition in TE mode as a function of temperature. As the thermally excited fermions begin to occupy higher energy states the E1--H1 transition weakens. A physical equivalent of the TME is the oscillator strength $f_{cv}\propto\mu_{cv}^2$, where $\mu_{cv}$ is the transition dipole moment, and it can be calculated from the TME using $\mu_{cv}=e\hbar\mathcal{P}_{cv}/im_0E_{cv}$. It allows us to quantify the transition strength from state $\ket{c}$ to $\ket{v}$, which decreases with a rise in temperature.

Now we shall study the spatial manifestation of the electronic states. Fig.\ \ref{fig:wfs-cdse22_8_4_300} shows the spatial charge densities of the first ten electron and hole states (square of the wavefunction $\vert\psi^2\vert$, i.e. probability of finding them), within the vertically central \textit{x-y} plane of the 4 ML CdSe NPL at 30$^\circ$C corresponding to Fig.\ \ref{fig:EN_prob-tempr}a. The charge density description in terms of \textit{s} and \textit{p} orbitals in a quite general feature of III-V semiconductors. However in II-VI semiconductors the VB is influenced by chemically active \textit{d} orbitals also.\cite{wei88} The E1, H1 and H7 states are $s$-like. While H1 is dominated by \textit{hh}, H7 has higher \textit{lh} contribution. The E2 and H2 are $p_y$-like, while E4, H4 and H9 are $p_x$-like -- H4 has greater \textit{hh} influence while H9 has more of \textit{lh}. The E3, E7 and H3 are formed by \textit{s-p}-mixing. The E5 and H5 are $d_{xy}$-like, while the E6 and H6 again have some amount of mixing. It is observed that with changes in temperature the spatial charge densities have very inconsequential variations because it is a measure of the electron and hole probability density determined by the wavefunction localization, which is primarily affected by piezoelectric strain and external electric fields.

       		\begin{figure}[t]
			\centering
			\includegraphics[scale=0.5]{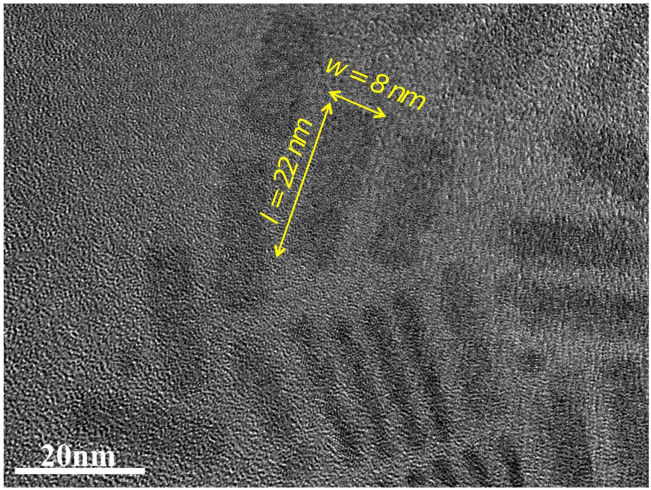}
			\caption{High-resolution transmission electron microscopy (TEM) image of 4 ML CdSe NPL ensemble population. Average lateral dimension is $l=22$ nm and $w=8$ nm (Scale: 20 nm)}
		\label{fig:4ML_cdse_NPL_TEM}
		\end{figure}

 	\begin{figure*}[t]
		\centering
		\includegraphics[width=1\textwidth]{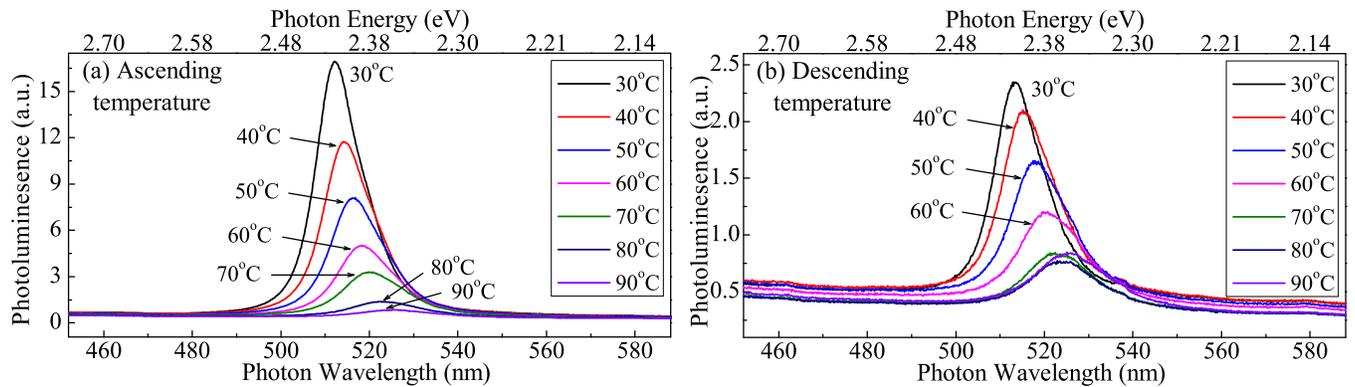}
		\caption{PL spectra of 4 ML CdSe NPLs measured in thin-film form while (a: \textit{left}) ascending, and (b: \textit{right}) descending temperature, respectively. Temperature of each PL spectrum is indicated. With an increase in temperature there is a redshift in the emission peak, broadening in the PL linewidth and reduction in PL intensity. Upon decreasing the temperature to RT, we observe substantial retractability in the peak position and linewidth, but not in the intensity. Fig.\ \ref{fig:EN-lw-Int-vs-T} shows the PL emission energy (peak position), linewidth and integrated intensity as a function of temperature.}
		\label{fig:pl-temp_inc_dec}
	\end{figure*}

	\begin{figure*}[t]
			\centering
			\includegraphics[width=1\textwidth]{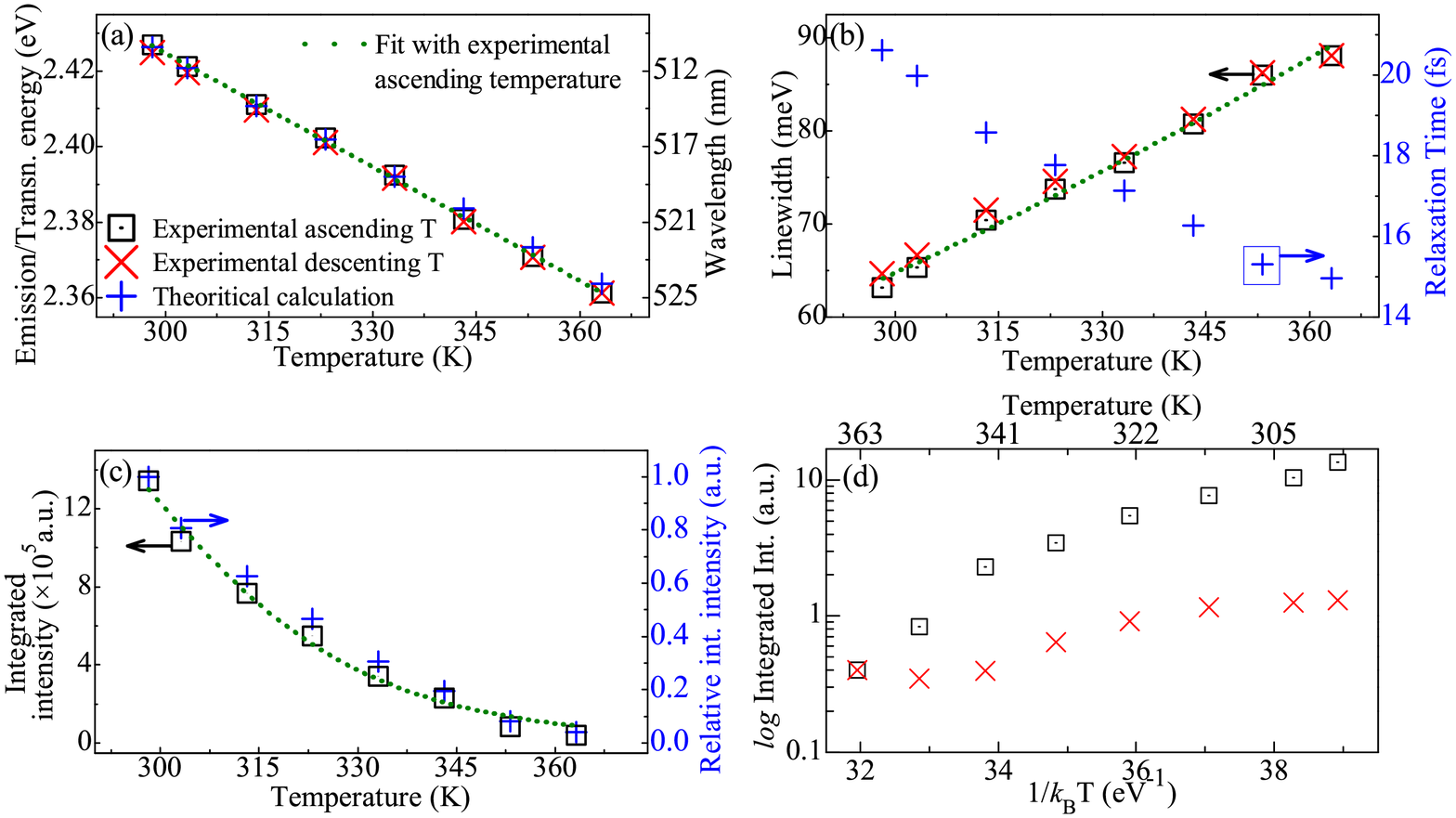}
			\caption{(a: \textit{top left}) Experimental PL emission energy/wavelength while ascending and descending temperature compared with theoretically simulated E1--H1 transition energy values; (b: \textit{top right}) Experimental PL linewidth while ascending and descending temperature and extracted intraband relaxation time $\tau_{in}$ used in theoretical model [Eq.\ \ref{eq:lorentz}] from data of ascending temperature; (c: \textit{bottom left}) Experimental PL intensity while ascending temperature compared with theoretically simulated relative PL integrated intensity; and (d: \textit{bottom right}) Comparative Arrhenius plot for PL integrated intensity (log scale) while ascending and descending temperature. Consistent symbols have been used in all the frames: \textbf{black $\boxdot$} and \textcolor{red}{\textbf{red $\times$}} for experimental data while ascending and descending temperature respectively, while \textcolor{blue}{\textbf{blue +}} for theoretically predicted data or data used in theoretical model (comparison is discussed in Sec.\ \ref{subsec:comp-theor-expt-res}). The \textcolor{ForestGreen}{\textbf{green $\cdots$}} dotted lines in frames (a), (b) and (c) show the fitted plots [using Eq.\ \ref{eqn:cardonaEq}, \ref{eq:linewidthEq} and \ref{eq:Int-T}] with the data of ascending temperature.}
			\label{fig:EN-lw-Int-vs-T}
	\end{figure*}

\subsection{\label{subsec:expt-res}Experimental Results}

We have synthesized 4 monolayer (ML) CdSe NPLs \textcolor{black}{as they have been most widely investigated in literature\cite{bose16,olutas16,bertrand16,guzelturk15,Achtstein15_jpcc,tessier14,Mahler12_jacs} and found applications in LED color-conversion layers, lasers and luminescent solar concentrators. We have used colloidal synthesis techniques similar to those used in Refs.\ \citen{Kelestemur_AFM16,tessier14,olutas15,guzelturk_acsNano14}, as described in the \textit{Experimental Methods} section.}

Fig.\ \ref{fig:4ML_cdse_NPL_TEM} shows a high-resolution transmission electron microscopy (TEM) image of our synthesized NPL ensemble population. We can observe NPLs in two orientations: face view (\textcolor{black}{laterally} flat) and edge view (\textcolor{black}{vertically} stacked). \textcolor{black}{We have done lateral dimension measurements over several NPL samples using ImageJ software. Their average length and width were measured to be 22 nm and 8 nm respectively.} The theoretical results discussed in Sec.\ \ref{subsec:theor-res} correspond to the same dimensions of our synthesized NPLs. \textcolor{black}{Measurements of stacked NPLs reveal differing thicknesses at the two ends, which is attributed to NPL tilting/folding, a common phenomenon for NPLs, also observed by several other groups.\cite{ithurria11,Hutter14} Also, the ligands surrounding the NPLs have a contrast in the TEM image, so the NPL boundary is not strictly discernible. However, the confirmation of 4 ML (1.2 nm) thickness comes from the optical spectra characterization. This is discussed in the ancillary document.} 

Spectral characterization results in Fig.\ \ref{fig:PL-Abs-TRPL} show the experimental PL emission and absorption spectrum (both measured in solution form) for our 4 ML CdSe NPLs at 30$^\circ$C. The PL was measured in a standard PL spectrometer with an excitation wavelength of 350 nm. We can observe the PL emission peak at 512 nm, which is in very good agreement with the E1--H1 transition energy prediction of 2.42 eV by our model, as shown in Fig.\ \ref{fig:EN_prob-tempr}a--b. Similarly, with our E1--H1 predictions at other varying temperatures (Fig.\ \ref{fig:EN_prob-tempr}a) we can determine their expected PL emission wavelengths.

With a Stokes shift of 3 nm from the PL spectrum, the primary absorption peak occurs at 509 nm. This is due to the reorganization of the solvent in solution-processed NPLs plus dissipation and vibrational relaxation of the NPLs.\cite{Lakowicz-book06} The primary absorption peak corresponds to the E1--H1 transition, where the H1 state is 2D-heavy holes (\textit{hh}) dominated [as theoretically determined in Fig.\ \ref{fig:EN_prob-tempr}b]. A study of the TMEs make this clear. The transverse electric (TE) mode TME for the E1--H1 transition is maximum among all possible cases accounting to 0.455. The secondary absorption peak occurs at 480 nm. A study of the transverse magnetic (TM) mode TMEs of the transitions near the secondary absorption energy shows that this comes from E1-H9 transitions, for which the TME value is 0.358, which is weaker than the E1-H1 TME but is stronger than TMEs of other allowed transitions. The fraction of light holes (\textit{lh}) involved in this transition is much higher (66 \%) than their \textit{hh} counterpart (22 \%) [Fig.\ \ref{fig:EN_prob-tempr}b]. This conforms with the idea that \textit{hh} contributes to the primary absorption peak of a typical CdSe NPL, while \textit{lh} contributes to the secondary peak.\cite{achtstein12} We have discussed about the TE and TM mode TME values for all possible transitions between E$_{\text{1}-\text{10}}$ to H$_{\text{1}-\text{10}}$ in the ancillary document. \textcolor{black}{Note that, the PL peak position at 512 nm, and the \textit{hh} and \textit{lh} absorption peaks at 509 nm and 480 nm respectively confirms that our synthesized NPL samples are indeed 4 ML (1.2 nm) thick.\cite{tessier14,Erdem16}}

Time-Resolved PL (TRPL) spectrum and fit at 30$^\circ$C (inset of Fig.\ \ref{fig:PL-Abs-TRPL}) reveals a dual decay path mechanism following the model $I(t)=a_1e^{-t/\tau_1}+a_2e^{-t/\tau_2}$, with PL decay lifetimes of $\tau_1=0.34$ ns and $\tau_2=2.27$ ns with 30\% and 70\% contribution respectively, leading to an average lifetime of 2.15 ns calculated using $\tau_{\text{avg}}=\left(a_1\tau_1^2+a_2\tau_2^2\right)/\left(a_1\tau_1+a_2\tau_2\right)$, which is similar to previously reported $\tau_{\text{avg}}$ of CdSe NPLs at RT.\cite{ithurria11,tessier13} The TRPL fit is shown again with the system response in the ancillary document. Previous radiative transition dynamics studies on CdSe NPLs have also observed a similar dual decay path mechanism where $\tau_1$ is assigned as the radiative recombination decay time period relating to reversible PL losses, while $\tau_2$ as the nonradiative recombination decay time period relating to the trap states which may cause irreversible losses in the PL intensity, \cite{achtstein12,halder14} originating from the fast hole-trapping due to incomplete passivation of Cd atoms on the NPL surface.\cite{Kelestemur_AFM16}

Fig.\ \ref{fig:pl-temp_inc_dec}a and Fig.\ \ref{fig:pl-temp_inc_dec}b shows the effect of temperature on the experimentally measured PL spectra of the NPLs in thin-film form using a laser excitation wavelength of 355 nm at 0.5 mW power, while ascending and descending temperature respectively. For every 10$^\circ$C rise in temperature, there is a red-shift in the emission wavelength of $\sim$2 nm, and the PL intensity falls as the sample quality degrades. Also, the spectrum broadens due to intraband scattering effect. Fig.\ \ref{fig:EN-lw-Int-vs-T} shows the photon emission energy, PL linewidth and PL integrated intensity as a function of ascending (black \textcolor{black}{$\boxdot$}) and descending (\textcolor{red}{red $\times$}) temperature. For a particular temperature the emission energy and linewidth are almost identical in both. While cooling back to RT, the signal intensity increases, but is far from being retraced. The recovery is $<14$ \% (see logarithmic comparison in inset of Fig.\ \ref{fig:EN-lw-Int-vs-T}c). The effective photon emission energy (equivalent to the effective E1-H1 transition energy) has a negative thermal coefficient as already discussed in the context of Fig.\ \ref{fig:EN_prob-tempr}a. For II-VI semiconductors, a semiempirical expression (Eq.\ \ref{eqn:cardonaEq}) suggested by Cardona \textit{et al.} \cite{logothetidis86} takes into consideration the phonon emission and absorption using Bose-Einstein statistical factors.

	\begin{equation}	E_{\text{exc}}\left(T\right)=E_{\text{exc}}\left(0\right)-a_{ep}\left(2n_B+1\right)
	\label{eqn:cardonaEq}
	\end{equation} 
    
Here $a_{ep}$ is the exciton-phonon coupling constant and $n_B=1/\left(e^{\Theta/T}-1\right)$. $\Theta$ is the average phonon temperature, such that the average phonon energy is $k_B\Theta$. Both, acoustic and optical phonons contribute to the redshift incurred by increasing temperature. $E_{\text{exc}}\left(0\right)$ is the exciton energy at 0 K. $T$ is the absolute temperature. Upon fitting the ascending temperature data of Fig. \ref{fig:EN-lw-Int-vs-T}a with Eq. \ref{eqn:cardonaEq}, we get $E_{\text{exc}}\left(0\right)=2.727$ eV. This value is in good agreement with results of Achtstein \textit{et al.},\cite{achtstein12} who reported 2.709 eV. This elevated $E_{\text{exc}}\left(0\right)$ compared to bulk like ZB CdSe epilayers (1.739 eV) \cite{chia08} is due to increased quantum confinement effects. Further, we obtain $a_{ep}=18.2\pm2.8$ meV and $\Theta=36\pm0.4$ K. Chia \textit{et al.}\cite{chia08} calculated $a_{ep}=21$ meV for bulk like ZB CdSe epilayers. This shows that the increased confinement in the NPLs has reduced the average exciton-phonon coupling, also evident from the significant reduction in average phonon temperature of 305 K in bulk ZB CdSe.\cite{madelung}

Moreover, the exciton-phonon interaction induced intraband scattering causes transition energy broadening. The temperature dependence of the linewidth of excitonic peaks is given by \cite{rudin90,Chuang}
	\begin{equation}
	\Gamma\left(T\right)=\Gamma_0+\gamma_{AC}T+\dfrac{\Gamma_{LO}}{e^{\left(\hbar\omega_{LO}/k_BT\right)}-1}
	\label{eq:linewidthEq}
	\end{equation}

\noindent where $\Gamma_0$ is the inhomogeneous broadening due to intrinsic effects (eg. alloy disorder, impurities). $\gamma_{AC}$ accounts for the exciton-acoustical phonon interaction. The last term comes from exciton-longitudinal optical (LO) phonon Fr\"{o}hlich interaction. For practical purposes, the contribution of $\gamma_{AC}$ compared to $\Gamma_{LO}$ is negligible, hence sometimes it is considered zero. \cite{achtstein12,perna98} At cryogenic range, particularly $<50$ K, the linewidth gradient drastically decreases. \cite{achtstein12,logothetidis86} therefore extrapolating our data of $\ge$ RT to lower temperatures would underestimate $\Gamma_0$. We have used $\Gamma_0=32.5$ meV from ref. \citen{achtstein12} following their linear fit at a low temperature range of 10-50 K. Further, based on the Raman spectra measurements of CdSe NPLs \cite{achtstein12} and QDs \cite{jing09}, we have taken $\hbar\omega_{LO}=25$ meV, to fit the other parameters of Eq. \ref{eq:linewidthEq} using the ascending temperature data from Fig.\ \ref{fig:EN-lw-Int-vs-T}b. We found $\Gamma_{LO}=236.3\pm11.8$ meV and $\gamma_{AC}=0.38\pm0.03$ meV, suggesting that the LO phonon contribution is much larger compared to that of the acoustic phonon, consistent with the calculations of Rudin \textit{et al.}\cite{rudin90} Comparing with bulk-like CdSe epilayers ($\gamma_{AC}=0.084$ meV)\cite{chia08}, NPLs have much higher $\gamma_{AC}$ due to an increase in acoustic phonon coupling caused by reduced dimensionality of the system, which is theoretically consistent \cite{takagahara93}. The interaction between excitons and LO phonons is dictated by the Fr\"{o}hlich interaction, which is Coulombic in nature caused by the longitudinal electric field produced by LO phonons. Therefore, $\hbar\omega_{LO}$ and $\Gamma_{LO}$ are related in the sense that a larger $\hbar\omega_{LO}$ induces a stronger Fr\"{o}hlich interaction and a larger $\Gamma_{LO}$. Valerini \textit{et al.} \cite{valerini05} have shown the presence of multiple temperature-dependent non-radiative processes in PL relaxation dynamics. At higher temperature, the contribution of LO phonons becomes predominant, eventually dominating the excitonic linewidth. \cite{perna98}

Furthermore, with increasing temperature, we also have a substantial reduction in the PL intensity as the excitons are trapped in the nonradiative centers, likely at NPL surfaces and dissociate into the continuum states. The integrated intensity of the PL spectra with varying temperature is given by \cite{chia08,leroux99}
	\begin{equation}	I\left(T\right)=I\left(0\right)/\left[1+C\cdot e^{-E_a/k_BT}\right]
	\label{eq:Int-T}
	\end{equation}       
\noindent where $I\left(0\right)$ and $I\left(T\right)$ are the integrated PL intensities at 0 K and \textit{T} K, respectively. The fitting constant $C$ is related to the radiative and nonradiative lifetimes, $E_a$ is the activation energy of the nonradiative channel. We fit the ascending temperature data of Fig.\ \ref{fig:EN-lw-Int-vs-T}c with Eq. \ref{eq:Int-T} to obtain $E_a=494\pm6$ meV. Heating above RT induces both reversible and irreversible losses in the intensity, and the latter restricts it to trace back upon cooling \cite{cai13} (inset in Fig.\ \ref{fig:EN-lw-Int-vs-T}c). Irreversible losses are generally caused by chemical processes like ligand loss or oxidative degradation damaging the NPL surface which disturbs the repopulation of mobile holes and electrons.\cite{liu06} The extent to which the PL intensity is retraced in Fig.\ \ref{fig:pl-temp_inc_dec}b is due to reversible losses, caused due to dynamic and static quenching.\cite{cai13} Upon heating, dynamic quenching activates the thermally activated non-radiative recombination paths compared to radiative paths; while static quenching causes the ratio of dark to bright particles to increase inducing non-radiative decay. While actively heating up to 90$^\circ$C took us a few minutes, every successive 10$^\circ$C fall took increasingly longer time -- 2, 4, 6, 11, 19 and 32 minutes till 30$^\circ$C and even longer to RT. The Newton's cooling constant following $T(t)=T_a+\Delta T_0e^{-kt}$ yeilds $k=0.089$ min\textsuperscript{-1}. An attempt to fit the descending \textit{T} integrated intensity with Eq. \ref{eq:Int-T} yeilds a poor Chi-square fit, as expected, since the irreversible losses incurred over almost a span of 1 hr disturbs the liable excitonic properties. If the cooling is conducted in a vacuum environment (to avoid losses due to oxidation), it will result in an improved intensity recovery.    
        
\subsection{\label{subsec:comp-theor-expt-res}Comparison of Theoretical and Experimental Results}

		\begin{figure}[t]
			\centering
			\includegraphics[trim={1.05cm 0cm 1.45cm 0.4cm},width=3.3in,height=1.85in]{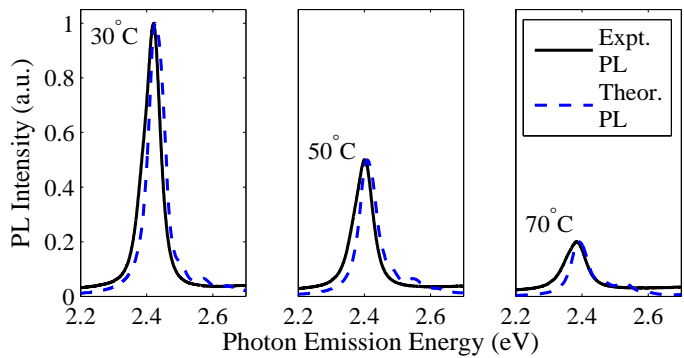}
			\caption{Comparison of experimental (solid black) vs. theoretical (dotted blue) PL spectra of 4 ML CdSe NPLs at 30, 50 and 70$^\circ$C.}
			\label{fig:sim-pl}
		\end{figure}
        
In Sec.\ \ref{subsec:theor-res} and \ref{subsec:expt-res}, we have studied the temperature-dependent optoelectronic properties of CdSe NPLs from a theoretical and experimental point-of-view. Here we combine and compare our findings, and simulate the PL spectra for varying temperature, as shown in Fig.\ \ref{fig:sim-pl}. A PL spectrum is characterized by three main parameters:
\begin{itemize}
  \item \textit{PL peak position}: We have theoretically calculated the temperature-dependent E1--H1 transition energy values (Fig.\ \ref{fig:EN_prob-tempr}a), \textcolor{black}{using the 8-band \textbf{\textit{k$\cdot$p}} method described in Sec.\ \ref{subsec:elec-bndstr} and material parameters from Table \ref{tab:mat-param}. The Hamiltonian matrix was solved by direct diagonalization to obtain the electronic structure and eigenenergy values, and E1--H1 transition energy (equivalent of the PL peak position).} It match excellently with the experimental data of PL peak position, as shown in Fig.\ \ref{fig:EN-lw-Int-vs-T}a.
\item \textit{PL linewidth}: The PL spectrum undergoes spectral broadening due to several empirical factors, such as local and extended structural defects, size dispersion, alloy disorders and impurities, which affects its linewidth.\cite{calderon} Thus, different 4 ML CdSe NPL samples may have varying linewidths at the same temperature. We have used the experimental linewidth data of ascending temperature to extract the intraband relaxation time ($\tau_{in}=2\hbar$/FWHM) used in our simulation \textcolor{black}{as described in the calculation of spontaneous radiative rate from the excitonic bound states and continuum-states in Sec.\ \ref{subsec:opt-prop}. This helps to model the linewidth of the simulated PL spectra according to the Lorentzian broadening lineshape [Eq.\ \ref{eq:lorentz}] for an accurate estimate.} The extracted $\tau_{in}$ is shown in Fig.\ \ref{fig:EN-lw-Int-vs-T}b.
\item \textit{PL intensity}: We have calculated the relative PL intensity, which shows an exponential fall with temperature, as verified experimentally. However, the experimental PL intensity (arbitrary units) is not the absolute total emission. It is therefore of interest to study and compare the relative intensities variation with temperature, which is influenced by several parameters such as Fermi factor, TME, etc. [see Eq.\ \ref{eq:ex} and \ref{eq:c}]. Our theoretically predicted relative PL intensities have a very good comparison with the experimental intensities of ascending temperature, as shown in Fig.\ \ref{fig:EN-lw-Int-vs-T}c.
\end{itemize}

The database of theoretical and experimental (ascending and descending temperature) results is given in the ancillary document. Employing the above mentioned approach, we have studied the PL spectra of the 4 ML CdSe NPLs at varying temperature. Fig.\ \ref{fig:sim-pl} shows a comparison between the experimental and theoretically calculated PL spectra at 30$^\circ$C, 50$^\circ$C and 70$^\circ$C. For a clear comparison of peak positions, linewidths and intensities, we have normalized both, the experimental and theoretical PL spectra with respect to the PL peak at RT. The experimental and theoretical PL spectra match appreciably, consistent with the results presented in Fig.\ \ref{fig:EN-lw-Int-vs-T}. In the ancillary document, we have compared, extended and validated our calculations and measurements with existing low temperature results from literature. We have studied the temperature-dependent PL peak positions (E1--H1 transition energies), PL linewidth and relative PL intensity in tandem with results from seminal works of Achtstein \textit{et al.}\cite{achtstein12} and Erdem \textit{et al.}\cite{Erdem16} to validate our results.
  
\section{Summary and Conclusion}
In summary, we have comprehensively studied the electronic bandstructure, probability in band mixing, charge densities, optical
transition matrix element, characteristic PL emission energy, linewidth and integrated intensity of CdSe NPLs -- as a function of temperature, which is of critical importance particularly for commercial device applications at elevated temperature. Our theoretical investigations are based on a framework that relies on the effective-mass envelope function theory and density-matrix theory. From the quantum physics and optoelectronics point-of-view, the implications of rise in operation temperature of NPLs are many-fold, such as (\textit{i}) reduction in effective band-edge transition energy, (\textit{ii}) fall in optical transition matrix element strength, (\textit{iii}) reduction in intraband state gaps, faster so in CB compared to VB, (\textit{iv}) promotion of interband \textit{e-h} coupling, (\textit{v}) modification in \textit{hh}-\textit{lh}-\textit{so} band mixing probabilities,
(\textit{vi}) reduction in Fermi factor, (\textit{vii}) shift of CB and VB quasi Fermi energy levels towards the band edges, and reduction in quasi Fermi separation, (\textit{viii}) invariance in the charge density profile, (\textit{ix}) redshift in peak photon emission energy in PL spectra, (\textit{x}) widening of PL spectral linewidth owing to intraband scattering, (\textit{xi}) exponential fall in PL integrated intensity, and (\textit{xii}) phenomenological evidence of reversible and irreversible losses induced which affects the sample quality and dictates the PL traceability.

In conclusion, we have theoretically and experimentally studied the temperature-dependent optoelectronic characteristics of quasi-2D colloidal CdSe NPLs and established a model to predict and study the electronic bandstructure and PL spectra of NPLs at any arbitrary temperature, which can be effectively used by experimentalists to optimize device design iterations.

\section*{Experimental Methods}
\subsection*{Synthesis Techniques}
We have synthesized 4 ML colloidal CdSe NPLs using synthesis protocol similar to those used in Refs.\ \citen{Kelestemur_AFM16,tessier14,olutas15,guzelturk_acsNano14}. The following chemicals were used: Cadmium acetate dihydrate Cd(OAc)$_2\cdot$2H$_2$O, oleic acid, technical grade 1-octadecene, cadmium nitrate, cadmium nitrate, \textit{n}-hexane, powder form selenium and myristic acid sodium salt. These were procured from Sigma Aldrich. The first step is to synthesize Cadmium (myristate)$_2$, by dissolving 1.23 g of Cadmium nitrate in methanol (40 mL). This was followed by dissolving sodium myristate (3.13 g) in methanol (250 mL) for 60 min with continuous stirring. When the two solutions were mixed, a white precipitate is formed after full dissolution. This precipitate was washed with methanol and dried under vacuum for about 24 h. 12 mg (0.15 mmol) of selenium (Se) and 170 mg (0.3 nmol) of Cd(myristate)$_2$ were introduced in a three-neck flask along with 15 mL of octadecene (ODE) and degassed under vacuum. Under argon (Ar) flow, the mixture was heated up, until 240$^\circ$C, and left there for 10 mins. When the color of the solution becomes yellowish, at $\sim195^\circ$C, 80 mg (0.30 mmol) of Cd(OAc)$_2\cdot$2H$_2$O was introduced. 5 mL of \textit{n}-hexane and 0.5 mL of oleic acid was added at the end of the synthesis, and the solution was allowed to cool down to RT. A centrifuge was used to precipitate the NPLs before storing them in hexane. This synthesis also produces some 5 ML NPLs and spherical nanocrystals with distinctly different emission wavelength. The 4 ML NPLs were separated from the rest using selective precipitation.

\subsection*{Purification and Film preparation}
The synthesized NPLs were centrifuged for 5 min at 4500 rpm, followed by removal of the supernatant solution. Nitrogen was used to dry the precipitate, before dissolving it in hexane to centrifuge for 10 min at 4500 rpm. The supernatant solution was made turbid by addition of ethanol, followed by centrifuging again for 10 min at 4500 rpm. Finally, using a 0.20 $\mu$m filter, the dissolved precipitate in hexane was filtered. Later, piranha solution was used for 30 min to clean quartz substrates of size 1.5 $\times$ 1.5 cm followed by DI-water. It was dried at 80$^\circ$C for 30 min in an oven and the filtered solutions was spin-coated for 1 min at 2000 rpm to prepare solid thin-films of NPLs.
        
\subsection*{\textcolor{black}{TEM and }Optical Characterization}
High-resolution transmission electron microscopy (TEM) images of NPL samples were taken on a JEOL JEM-2010 HR TEM operating at 200 kV, \textcolor{black}{which has a resolution of $\sim$0.22 nm}. The steady state PL spectra measurements were done using a cooled iDus CCD system. The time resolved PL measurements were done with a Becker \& Hickl DCS 120 confocal scanning FLIM system with an excitation laser of 375 nm. The system has temporal resolution of 200 ps.

\begin{acknowledgments}
We acknowledge the Ministry of Education, Singapore (RG 182/14 and RG 86/13), A*STAR (1220703063), Economic Development Board (NRF2013SAS-SRP001-019) and Asian Office of Aerospace Research and Development (FA2386-17-1-0039).
\end{acknowledgments}

\bibliography{_my-refs}

\begin{thebibliography}{55}%
\makeatletter
\providecommand \@ifxundefined [1]{%
 \@ifx{#1\undefined}
}%
\providecommand \@ifnum [1]{%
 \ifnum #1\expandafter \@firstoftwo
 \else \expandafter \@secondoftwo
 \fi
}%
\providecommand \@ifx [1]{%
 \ifx #1\expandafter \@firstoftwo
 \else \expandafter \@secondoftwo
 \fi
}%
\providecommand \natexlab [1]{#1}%
\providecommand \enquote  [1]{``#1''}%
\providecommand \bibnamefont  [1]{#1}%
\providecommand \bibfnamefont [1]{#1}%
\providecommand \citenamefont [1]{#1}%
\providecommand \href@noop [0]{\@secondoftwo}%
\providecommand \href [0]{\begingroup \@sanitize@url \@href}%
\providecommand \@href[1]{\@@startlink{#1}\@@href}%
\providecommand \@@href[1]{\endgroup#1\@@endlink}%
\providecommand \@sanitize@url [0]{\catcode `\\12\catcode `\$12\catcode
  `\&12\catcode `\#12\catcode `\^12\catcode `\_12\catcode `\%12\relax}%
\providecommand \@@startlink[1]{}%
\providecommand \@@endlink[0]{}%
\providecommand \url  [0]{\begingroup\@sanitize@url \@url }%
\providecommand \@url [1]{\endgroup\@href {#1}{\urlprefix }}%
\providecommand \urlprefix  [0]{URL }%
\providecommand \Eprint [0]{\href }%
\providecommand \doibase [0]{http://dx.doi.org/}%
\providecommand \selectlanguage [0]{\@gobble}%
\providecommand \bibinfo  [0]{\@secondoftwo}%
\providecommand \bibfield  [0]{\@secondoftwo}%
\providecommand \translation [1]{[#1]}%
\providecommand \BibitemOpen [0]{}%
\providecommand \bibitemStop [0]{}%
\providecommand \bibitemNoStop [0]{.\EOS\space}%
\providecommand \EOS [0]{\spacefactor3000\relax}%
\providecommand \BibitemShut  [1]{\csname bibitem#1\endcsname}%
\let\auto@bib@innerbib\@empty
\bibitem [{\citenamefont {Ithurria}\ and\ \citenamefont
  {Dubertret}(2008)}]{ithurria08}%
  \BibitemOpen
  \bibfield  {author} {\bibinfo {author} {\bibfnamefont {S.}~\bibnamefont
  {Ithurria}}\ and\ \bibinfo {author} {\bibfnamefont {B.}~\bibnamefont
  {Dubertret}},\ }\href {\doibase 10.1021/ja807724e} {\bibfield  {journal}
  {\bibinfo  {journal} {J. Am. Chem. Soc.}\ }\textbf {\bibinfo {volume}
  {130}},\ \bibinfo {pages} {16504} (\bibinfo {year} {2008})}\BibitemShut
  {NoStop}%
\bibitem [{\citenamefont {Guzelturk}\ \emph
  {et~al.}(2014{\natexlab{a}})\citenamefont {Guzelturk}, \citenamefont
  {Kelestemur}, \citenamefont {Olutas}, \citenamefont {Delikanli},\ and\
  \citenamefont {Demir}}]{guzelturk14}%
  \BibitemOpen
  \bibfield  {author} {\bibinfo {author} {\bibfnamefont {B.}~\bibnamefont
  {Guzelturk}}, \bibinfo {author} {\bibfnamefont {Y.}~\bibnamefont
  {Kelestemur}}, \bibinfo {author} {\bibfnamefont {M.}~\bibnamefont {Olutas}},
  \bibinfo {author} {\bibfnamefont {S.}~\bibnamefont {Delikanli}}, \ and\
  \bibinfo {author} {\bibfnamefont {H.~V.}\ \bibnamefont {Demir}},\ }\href
  {http://dx.doi.org/10.1021/nn5022296} {\bibfield  {journal} {\bibinfo
  {journal} {ACS Nano}\ }\textbf {\bibinfo {volume} {8}},\ \bibinfo {pages}
  {6599} (\bibinfo {year} {2014}{\natexlab{a}})}\BibitemShut {NoStop}%
\bibitem [{\citenamefont {Chen}\ \emph {et~al.}(2014)\citenamefont {Chen},
  \citenamefont {Nadal}, \citenamefont {Mahler}, \citenamefont {Aubin},\ and\
  \citenamefont {Dubertret}}]{chen14}%
  \BibitemOpen
  \bibfield  {author} {\bibinfo {author} {\bibfnamefont {Z.}~\bibnamefont
  {Chen}}, \bibinfo {author} {\bibfnamefont {B.}~\bibnamefont {Nadal}},
  \bibinfo {author} {\bibfnamefont {B.}~\bibnamefont {Mahler}}, \bibinfo
  {author} {\bibfnamefont {H.}~\bibnamefont {Aubin}}, \ and\ \bibinfo {author}
  {\bibfnamefont {B.}~\bibnamefont {Dubertret}},\ }\href {\doibase
  10.1002/adfm.201301711} {\bibfield  {journal} {\bibinfo  {journal} {Adv.
  Funct. Mater.}\ }\textbf {\bibinfo {volume} {24}},\ \bibinfo {pages} {295}
  (\bibinfo {year} {2014})}\BibitemShut {NoStop}%
\bibitem [{\citenamefont {Ithurria}\ \emph {et~al.}(2011)\citenamefont
  {Ithurria}, \citenamefont {Tessier}, \citenamefont {Mahler}, \citenamefont
  {Lobo}, \citenamefont {Dubertret},\ and\ \citenamefont {Efros}}]{ithurria11}%
  \BibitemOpen
  \bibfield  {author} {\bibinfo {author} {\bibfnamefont {S.}~\bibnamefont
  {Ithurria}}, \bibinfo {author} {\bibfnamefont {M.~D.}\ \bibnamefont
  {Tessier}}, \bibinfo {author} {\bibfnamefont {B.}~\bibnamefont {Mahler}},
  \bibinfo {author} {\bibfnamefont {R.~P.}\ \bibnamefont {Lobo}}, \bibinfo
  {author} {\bibfnamefont {B.}~\bibnamefont {Dubertret}}, \ and\ \bibinfo
  {author} {\bibfnamefont {A.~L.}\ \bibnamefont {Efros}},\ }\href {\doibase
  10.1038/nmat3145} {\bibfield  {journal} {\bibinfo  {journal} {Nat. Mater.}\
  }\textbf {\bibinfo {volume} {10}},\ \bibinfo {pages} {936} (\bibinfo {year}
  {2011})}\BibitemShut {NoStop}%
\bibitem [{\citenamefont {Benchamekh}\ \emph {et~al.}(2014)\citenamefont
  {Benchamekh}, \citenamefont {Gippius}, \citenamefont {Even}, \citenamefont
  {Nestoklon}, \citenamefont {Jancu}, \citenamefont {Ithurria}, \citenamefont
  {Dubertret}, \citenamefont {Efros},\ and\ \citenamefont
  {Voisin}}]{benchamekh14}%
  \BibitemOpen
  \bibfield  {author} {\bibinfo {author} {\bibfnamefont {R.}~\bibnamefont
  {Benchamekh}}, \bibinfo {author} {\bibfnamefont {N.~A.}\ \bibnamefont
  {Gippius}}, \bibinfo {author} {\bibfnamefont {J.}~\bibnamefont {Even}},
  \bibinfo {author} {\bibfnamefont {M.~O.}\ \bibnamefont {Nestoklon}}, \bibinfo
  {author} {\bibfnamefont {J.~M.}\ \bibnamefont {Jancu}}, \bibinfo {author}
  {\bibfnamefont {S.}~\bibnamefont {Ithurria}}, \bibinfo {author}
  {\bibfnamefont {B.}~\bibnamefont {Dubertret}}, \bibinfo {author}
  {\bibfnamefont {A.~L.}\ \bibnamefont {Efros}}, \ and\ \bibinfo {author}
  {\bibfnamefont {P.}~\bibnamefont {Voisin}},\ }\href@noop {} {\bibfield
  {journal} {\bibinfo  {journal} {Phys. Rev. B}\ }\textbf {\bibinfo {volume}
  {89}},\ \bibinfo {pages} {035307} (\bibinfo {year} {2014})}\BibitemShut
  {NoStop}%
\bibitem [{\citenamefont {Tessier}\ \emph {et~al.}(2013)\citenamefont
  {Tessier}, \citenamefont {Mahler}, \citenamefont {Nadal}, \citenamefont
  {Heuclin}, \citenamefont {Pedetti},\ and\ \citenamefont
  {Dubertret}}]{tessier13}%
  \BibitemOpen
  \bibfield  {author} {\bibinfo {author} {\bibfnamefont {M.~D.}\ \bibnamefont
  {Tessier}}, \bibinfo {author} {\bibfnamefont {B.}~\bibnamefont {Mahler}},
  \bibinfo {author} {\bibfnamefont {B.}~\bibnamefont {Nadal}}, \bibinfo
  {author} {\bibfnamefont {H.}~\bibnamefont {Heuclin}}, \bibinfo {author}
  {\bibfnamefont {S.}~\bibnamefont {Pedetti}}, \ and\ \bibinfo {author}
  {\bibfnamefont {B.}~\bibnamefont {Dubertret}},\ }\href {\doibase
  10.1021/nl401538n} {\bibfield  {journal} {\bibinfo  {journal} {Nano Lett.}\
  }\textbf {\bibinfo {volume} {13}},\ \bibinfo {pages} {3321} (\bibinfo {year}
  {2013})}\BibitemShut {NoStop}%
\bibitem [{\citenamefont {Guzelturk}\ \emph {et~al.}(2015)\citenamefont
  {Guzelturk}, \citenamefont {Olutas}, \citenamefont {Delikanli}, \citenamefont
  {Kelestemur}, \citenamefont {Erdem},\ and\ \citenamefont
  {Demir}}]{guzelturk15}%
  \BibitemOpen
  \bibfield  {author} {\bibinfo {author} {\bibfnamefont {B.}~\bibnamefont
  {Guzelturk}}, \bibinfo {author} {\bibfnamefont {M.}~\bibnamefont {Olutas}},
  \bibinfo {author} {\bibfnamefont {S.}~\bibnamefont {Delikanli}}, \bibinfo
  {author} {\bibfnamefont {Y.}~\bibnamefont {Kelestemur}}, \bibinfo {author}
  {\bibfnamefont {O.}~\bibnamefont {Erdem}}, \ and\ \bibinfo {author}
  {\bibfnamefont {H.~V.}\ \bibnamefont {Demir}},\ }\href {\doibase
  10.1039/C4NR06003B} {\bibfield  {journal} {\bibinfo  {journal} {Nanoscale}\
  }\textbf {\bibinfo {volume} {7}},\ \bibinfo {pages} {2545} (\bibinfo {year}
  {2015})}\BibitemShut {NoStop}%
\bibitem [{\citenamefont {Grim}\ \emph {et~al.}(2014)\citenamefont {Grim},
  \citenamefont {Christodoulou}, \citenamefont {Di~Stasio}, \citenamefont
  {Krahne}, \citenamefont {Cingolani}, \citenamefont {Manna},\ and\
  \citenamefont {Moreels}}]{grim14}%
  \BibitemOpen
  \bibfield  {author} {\bibinfo {author} {\bibfnamefont {J.~Q.}\ \bibnamefont
  {Grim}}, \bibinfo {author} {\bibfnamefont {S.}~\bibnamefont {Christodoulou}},
  \bibinfo {author} {\bibfnamefont {F.}~\bibnamefont {Di~Stasio}}, \bibinfo
  {author} {\bibfnamefont {R.}~\bibnamefont {Krahne}}, \bibinfo {author}
  {\bibfnamefont {R.}~\bibnamefont {Cingolani}}, \bibinfo {author}
  {\bibfnamefont {L.}~\bibnamefont {Manna}}, \ and\ \bibinfo {author}
  {\bibfnamefont {I.}~\bibnamefont {Moreels}},\ }\href
  {http://dx.doi.org/10.1038/nnano.2014.213} {\bibfield  {journal} {\bibinfo
  {journal} {Nat. Nanotechnol.}\ }\textbf {\bibinfo {volume} {9}},\ \bibinfo
  {pages} {891} (\bibinfo {year} {2014})}\BibitemShut {NoStop}%
\bibitem [{\citenamefont {Zhao}\ \emph {et~al.}(2012)\citenamefont {Zhao},
  \citenamefont {Riemersma}, \citenamefont {Pietra}, \citenamefont {Koole},
  \citenamefont {de~Mello~Donegá},\ and\ \citenamefont {Meijerink}}]{zhao12b}%
  \BibitemOpen
  \bibfield  {author} {\bibinfo {author} {\bibfnamefont {Y.}~\bibnamefont
  {Zhao}}, \bibinfo {author} {\bibfnamefont {C.}~\bibnamefont {Riemersma}},
  \bibinfo {author} {\bibfnamefont {F.}~\bibnamefont {Pietra}}, \bibinfo
  {author} {\bibfnamefont {R.}~\bibnamefont {Koole}}, \bibinfo {author}
  {\bibfnamefont {C.}~\bibnamefont {de~Mello~Donegá}}, \ and\ \bibinfo
  {author} {\bibfnamefont {A.}~\bibnamefont {Meijerink}},\ }\href {\doibase
  10.1021/nn303217q} {\bibfield  {journal} {\bibinfo  {journal} {ACS Nano}\
  }\textbf {\bibinfo {volume} {6}},\ \bibinfo {pages} {9058} (\bibinfo {year}
  {2012})},\ \Eprint {http://arxiv.org/abs/http://dx.doi.org/10.1021/nn303217q}
  {http://dx.doi.org/10.1021/nn303217q} \BibitemShut {NoStop}%
\bibitem [{\citenamefont {Achtstein}\ \emph {et~al.}(2012)\citenamefont
  {Achtstein}, \citenamefont {Schliwa}, \citenamefont {Prudnikau},
  \citenamefont {Hardzei}, \citenamefont {Artemyev}, \citenamefont {Thomsen},\
  and\ \citenamefont {Woggon}}]{achtstein12}%
  \BibitemOpen
  \bibfield  {author} {\bibinfo {author} {\bibfnamefont {A.~W.}\ \bibnamefont
  {Achtstein}}, \bibinfo {author} {\bibfnamefont {A.}~\bibnamefont {Schliwa}},
  \bibinfo {author} {\bibfnamefont {A.}~\bibnamefont {Prudnikau}}, \bibinfo
  {author} {\bibfnamefont {M.}~\bibnamefont {Hardzei}}, \bibinfo {author}
  {\bibfnamefont {M.~V.}\ \bibnamefont {Artemyev}}, \bibinfo {author}
  {\bibfnamefont {C.}~\bibnamefont {Thomsen}}, \ and\ \bibinfo {author}
  {\bibfnamefont {U.}~\bibnamefont {Woggon}},\ }\href {\doibase
  10.1021/nl301071n} {\bibfield  {journal} {\bibinfo  {journal} {Nano Lett.}\
  }\textbf {\bibinfo {volume} {12}},\ \bibinfo {pages} {3151} (\bibinfo {year}
  {2012})}\BibitemShut {NoStop}%
\bibitem [{\citenamefont {Ji}\ \emph {et~al.}(2015)\citenamefont {Ji},
  \citenamefont {Zhang}, \citenamefont {Zhang}, \citenamefont {Liu},
  \citenamefont {Zhang}, \citenamefont {Shen}, \citenamefont {Wang},
  \citenamefont {Gao}, \citenamefont {Wang}, \citenamefont {Zhao},\ and\
  \citenamefont {Yu}}]{ji15}%
  \BibitemOpen
  \bibfield  {author} {\bibinfo {author} {\bibfnamefont {C.}~\bibnamefont
  {Ji}}, \bibinfo {author} {\bibfnamefont {Y.}~\bibnamefont {Zhang}}, \bibinfo
  {author} {\bibfnamefont {T.}~\bibnamefont {Zhang}}, \bibinfo {author}
  {\bibfnamefont {W.}~\bibnamefont {Liu}}, \bibinfo {author} {\bibfnamefont
  {X.}~\bibnamefont {Zhang}}, \bibinfo {author} {\bibfnamefont
  {H.}~\bibnamefont {Shen}}, \bibinfo {author} {\bibfnamefont {Y.}~\bibnamefont
  {Wang}}, \bibinfo {author} {\bibfnamefont {W.}~\bibnamefont {Gao}}, \bibinfo
  {author} {\bibfnamefont {Y.}~\bibnamefont {Wang}}, \bibinfo {author}
  {\bibfnamefont {J.}~\bibnamefont {Zhao}}, \ and\ \bibinfo {author}
  {\bibfnamefont {W.~W.}\ \bibnamefont {Yu}},\ }\href {\doibase
  10.1021/acs.jpcc.5b01030} {\bibfield  {journal} {\bibinfo  {journal} {J.
  Phys. Chem. C}\ }\textbf {\bibinfo {volume} {119}},\ \bibinfo {pages} {13841}
  (\bibinfo {year} {2015})},\ \Eprint
  {http://arxiv.org/abs/http://dx.doi.org/10.1021/acs.jpcc.5b01030}
  {http://dx.doi.org/10.1021/acs.jpcc.5b01030} \BibitemShut {NoStop}%
\bibitem [{\citenamefont {Jing}\ \emph {et~al.}(2009)\citenamefont {Jing},
  \citenamefont {Zheng}, \citenamefont {Ikezawa}, \citenamefont {Liu},
  \citenamefont {Lv}, \citenamefont {Kong}, \citenamefont {Zhao},\ and\
  \citenamefont {Masumoto}}]{jing09}%
  \BibitemOpen
  \bibfield  {author} {\bibinfo {author} {\bibfnamefont {P.}~\bibnamefont
  {Jing}}, \bibinfo {author} {\bibfnamefont {J.}~\bibnamefont {Zheng}},
  \bibinfo {author} {\bibfnamefont {M.}~\bibnamefont {Ikezawa}}, \bibinfo
  {author} {\bibfnamefont {X.}~\bibnamefont {Liu}}, \bibinfo {author}
  {\bibfnamefont {S.}~\bibnamefont {Lv}}, \bibinfo {author} {\bibfnamefont
  {X.}~\bibnamefont {Kong}}, \bibinfo {author} {\bibfnamefont {J.}~\bibnamefont
  {Zhao}}, \ and\ \bibinfo {author} {\bibfnamefont {Y.}~\bibnamefont
  {Masumoto}},\ }\href {http://dx.doi.org/10.1021/jp902080p} {\bibfield
  {journal} {\bibinfo  {journal} {J. Phys. Chem. C,}\ }\textbf {\bibinfo
  {volume} {113}},\ \bibinfo {pages} {13545} (\bibinfo {year}
  {2009})}\BibitemShut {NoStop}%
\bibitem [{\citenamefont {Le-Van}\ \emph {et~al.}(2015)\citenamefont {Le-Van},
  \citenamefont {Le~Roux}, \citenamefont {Teperik}, \citenamefont {Habert},
  \citenamefont {Marquier}, \citenamefont {Greffet},\ and\ \citenamefont
  {Degiron}}]{van15}%
  \BibitemOpen
  \bibfield  {author} {\bibinfo {author} {\bibfnamefont {Q.}~\bibnamefont
  {Le-Van}}, \bibinfo {author} {\bibfnamefont {X.}~\bibnamefont {Le~Roux}},
  \bibinfo {author} {\bibfnamefont {T.~V.}\ \bibnamefont {Teperik}}, \bibinfo
  {author} {\bibfnamefont {B.}~\bibnamefont {Habert}}, \bibinfo {author}
  {\bibfnamefont {F.}~\bibnamefont {Marquier}}, \bibinfo {author}
  {\bibfnamefont {J.-J.}\ \bibnamefont {Greffet}}, \ and\ \bibinfo {author}
  {\bibfnamefont {A.}~\bibnamefont {Degiron}},\ }\href {\doibase
  10.1103/PhysRevB.91.085412} {\bibfield  {journal} {\bibinfo  {journal} {Phys.
  Rev. B}\ }\textbf {\bibinfo {volume} {91}},\ \bibinfo {pages} {085412}
  (\bibinfo {year} {2015})}\BibitemShut {NoStop}%
\bibitem [{\citenamefont {Chen}\ \emph {et~al.}(2009)\citenamefont {Chen},
  \citenamefont {Fan}, \citenamefont {Xu}, \citenamefont {Zhang}, \citenamefont
  {Li},\ and\ \citenamefont {Xia}}]{chen09}%
  \BibitemOpen
  \bibfield  {author} {\bibinfo {author} {\bibfnamefont {J.}~\bibnamefont
  {Chen}}, \bibinfo {author} {\bibfnamefont {W.~J.}\ \bibnamefont {Fan}},
  \bibinfo {author} {\bibfnamefont {Q.}~\bibnamefont {Xu}}, \bibinfo {author}
  {\bibfnamefont {X.~W.}\ \bibnamefont {Zhang}}, \bibinfo {author}
  {\bibfnamefont {S.~S.}\ \bibnamefont {Li}}, \ and\ \bibinfo {author}
  {\bibfnamefont {J.~B.}\ \bibnamefont {Xia}},\ }\href {\doibase
  10.1063/1.3143025} {\bibfield  {journal} {\bibinfo  {journal} {J. Appl.
  Phys.}\ }\textbf {\bibinfo {volume} {105}},\ \bibinfo {pages} {123705}
  (\bibinfo {year} {2009})}\BibitemShut {NoStop}%
\bibitem [{\citenamefont {Song}\ \emph {et~al.}(2016)\citenamefont {Song},
  \citenamefont {Bose}, \citenamefont {Fan},\ and\ \citenamefont
  {Li}}]{song16}%
  \BibitemOpen
  \bibfield  {author} {\bibinfo {author} {\bibfnamefont {Z.-G.}\ \bibnamefont
  {Song}}, \bibinfo {author} {\bibfnamefont {S.}~\bibnamefont {Bose}}, \bibinfo
  {author} {\bibfnamefont {W.-J.}\ \bibnamefont {Fan}}, \ and\ \bibinfo
  {author} {\bibfnamefont {S.-S.}\ \bibnamefont {Li}},\ }\href {\doibase
  http://dx.doi.org/10.1063/1.4945700} {\bibfield  {journal} {\bibinfo
  {journal} {J. Appl. Phys.}\ }\textbf {\bibinfo {volume} {119}},\ \bibinfo
  {eid} {143103} (\bibinfo {year} {2016})}\BibitemShut {NoStop}%
\bibitem [{\citenamefont {Bose}\ \emph
  {et~al.}(2014{\natexlab{a}})\citenamefont {Bose}, \citenamefont {Fan},
  \citenamefont {Chen}, \citenamefont {Zhang},\ and\ \citenamefont
  {Tan}}]{bose14b}%
  \BibitemOpen
  \bibfield  {author} {\bibinfo {author} {\bibfnamefont {S.}~\bibnamefont
  {Bose}}, \bibinfo {author} {\bibfnamefont {W.~J.}\ \bibnamefont {Fan}},
  \bibinfo {author} {\bibfnamefont {J.}~\bibnamefont {Chen}}, \bibinfo {author}
  {\bibfnamefont {D.~H.}\ \bibnamefont {Zhang}}, \ and\ \bibinfo {author}
  {\bibfnamefont {C.~S.}\ \bibnamefont {Tan}},\ }in\ \href {\doibase
  10.1364/PHOTONICS.2014.S4D.5} {\emph {\bibinfo {booktitle} {12th
  International Conference on Fiber Optics and Photonics}}}\ (\bibinfo
  {publisher} {Optical Society of America},\ \bibinfo {year} {2014})\ p.\
  \bibinfo {pages} {S4D.5}\BibitemShut {NoStop}%
\bibitem [{\citenamefont {Bose}\ \emph
  {et~al.}(2014{\natexlab{b}})\citenamefont {Bose}, \citenamefont {Fan},
  \citenamefont {Jian}, \citenamefont {Zhang},\ and\ \citenamefont
  {Tan}}]{bose14a}%
  \BibitemOpen
  \bibfield  {author} {\bibinfo {author} {\bibfnamefont {S.}~\bibnamefont
  {Bose}}, \bibinfo {author} {\bibfnamefont {W.}~\bibnamefont {Fan}}, \bibinfo
  {author} {\bibfnamefont {C.}~\bibnamefont {Jian}}, \bibinfo {author}
  {\bibfnamefont {D.}~\bibnamefont {Zhang}}, \ and\ \bibinfo {author}
  {\bibfnamefont {C.}~\bibnamefont {Tan}},\ }in\ \href {\doibase
  10.1109/ISTDM.2014.6874705} {\emph {\bibinfo {booktitle} {Silicon-Germanium
  Technology and Device Meeting (ISTDM), 2014 7th International}}}\ (\bibinfo
  {year} {2014})\ pp.\ \bibinfo {pages} {129--130}\BibitemShut {NoStop}%
\bibitem [{\citenamefont {Bose}\ \emph {et~al.}(2016)\citenamefont {Bose},
  \citenamefont {Song}, \citenamefont {Fan},\ and\ \citenamefont
  {Zhang}}]{bose16}%
  \BibitemOpen
  \bibfield  {author} {\bibinfo {author} {\bibfnamefont {S.}~\bibnamefont
  {Bose}}, \bibinfo {author} {\bibfnamefont {Z.}~\bibnamefont {Song}}, \bibinfo
  {author} {\bibfnamefont {W.~J.}\ \bibnamefont {Fan}}, \ and\ \bibinfo
  {author} {\bibfnamefont {D.~H.}\ \bibnamefont {Zhang}},\ }\href
  {http://scitation.aip.org/content/aip/journal/jap/119/14/10.1063/1.4945993}
  {\bibfield  {journal} {\bibinfo  {journal} {J. Appl. Phys.}\ }\textbf
  {\bibinfo {volume} {119}},\ \bibinfo {pages} {143107} (\bibinfo {year}
  {2016})}\BibitemShut {NoStop}%
\bibitem [{\citenamefont {Shan}\ \emph {et~al.}(1994)\citenamefont {Shan},
  \citenamefont {Song}, \citenamefont {Luo},\ and\ \citenamefont
  {Furdyna}}]{shan94}%
  \BibitemOpen
  \bibfield  {author} {\bibinfo {author} {\bibfnamefont {W.}~\bibnamefont
  {Shan}}, \bibinfo {author} {\bibfnamefont {J.~J.}\ \bibnamefont {Song}},
  \bibinfo {author} {\bibfnamefont {H.}~\bibnamefont {Luo}}, \ and\ \bibinfo
  {author} {\bibfnamefont {J.~K.}\ \bibnamefont {Furdyna}},\ }\href
  {http://link.aps.org/doi/10.1103/PhysRevB.50.8012} {\bibfield  {journal}
  {\bibinfo  {journal} {Phys. Rev. B}\ }\textbf {\bibinfo {volume} {50}},\
  \bibinfo {pages} {8012} (\bibinfo {year} {1994})}\BibitemShut {NoStop}%
\bibitem [{\citenamefont {Karazhanov}\ and\ \citenamefont
  {Voon}(2005)}]{karazhanov05}%
  \BibitemOpen
  \bibfield  {author} {\bibinfo {author} {\bibfnamefont {S.~Z.}\ \bibnamefont
  {Karazhanov}}\ and\ \bibinfo {author} {\bibfnamefont {L.~L.~Y.}\ \bibnamefont
  {Voon}},\ }\href@noop {} {\bibfield  {journal} {\bibinfo  {journal}
  {Semicon.}\ }\textbf {\bibinfo {volume} {39}},\ \bibinfo {pages} {161}
  (\bibinfo {year} {2005})}\BibitemShut {NoStop}%
\bibitem [{\citenamefont {Madelung}(2004)}]{madelung}%
  \BibitemOpen
  \bibfield  {author} {\bibinfo {author} {\bibfnamefont {O.}~\bibnamefont
  {Madelung}},\ }\href@noop {} {\emph {\bibinfo {title} {{Semiconductors: Data
  Handbook}}}}\ (\bibinfo  {publisher} {Springer},\ \bibinfo {year}
  {2004})\BibitemShut {NoStop}%
\bibitem [{\citenamefont {Kumar}(2000)}]{kumar00}%
  \BibitemOpen
  \bibfield  {author} {\bibinfo {author} {\bibfnamefont {V.}~\bibnamefont
  {Kumar}},\ }\href {\doibase http://dx.doi.org/10.1016/S0022-3697(99)00238-3}
  {\bibfield  {journal} {\bibinfo  {journal} {J. Phys. Chem. of Solids}\
  }\textbf {\bibinfo {volume} {61}},\ \bibinfo {pages} {91 } (\bibinfo {year}
  {2000})}\BibitemShut {NoStop}%
\bibitem [{\citenamefont {Landry}\ \emph {et~al.}(2014)\citenamefont {Landry},
  \citenamefont {Morrell}, \citenamefont {Karagounis}, \citenamefont {Hsia},\
  and\ \citenamefont {Wang}}]{landry14}%
  \BibitemOpen
  \bibfield  {author} {\bibinfo {author} {\bibfnamefont {M.~L.}\ \bibnamefont
  {Landry}}, \bibinfo {author} {\bibfnamefont {T.~E.}\ \bibnamefont {Morrell}},
  \bibinfo {author} {\bibfnamefont {T.~K.}\ \bibnamefont {Karagounis}},
  \bibinfo {author} {\bibfnamefont {C.-H.}\ \bibnamefont {Hsia}}, \ and\
  \bibinfo {author} {\bibfnamefont {C.-Y.}\ \bibnamefont {Wang}},\ }\href
  {http://dx.doi.org/10.1021/ed300568e} {\bibfield  {journal} {\bibinfo
  {journal} {J. Chem. Educ.}\ }\textbf {\bibinfo {volume} {91}},\ \bibinfo
  {pages} {274} (\bibinfo {year} {2014})}\BibitemShut {NoStop}%
\bibitem [{\citenamefont {Micallef}, \citenamefont {Li},\ and\ \citenamefont
  {Weiss}(1993)}]{micallef93}%
  \BibitemOpen
  \bibfield  {author} {\bibinfo {author} {\bibfnamefont {J.}~\bibnamefont
  {Micallef}}, \bibinfo {author} {\bibfnamefont {E.}~\bibnamefont {Li}}, \ and\
  \bibinfo {author} {\bibfnamefont {B.~L.}\ \bibnamefont {Weiss}},\ }\href
  {\doibase http://dx.doi.org/10.1006/spmi.1993.1063} {\bibfield  {journal}
  {\bibinfo  {journal} {Superlattices Microstruct}\ }\textbf {\bibinfo {volume}
  {13}},\ \bibinfo {pages} {315} (\bibinfo {year} {1993})}\BibitemShut
  {NoStop}%
\bibitem [{\citenamefont {Li}\ and\ \citenamefont {Weiss}(1992)}]{herbert92}%
  \BibitemOpen
  \bibfield  {author} {\bibinfo {author} {\bibfnamefont {E.~H.}\ \bibnamefont
  {Li}}\ and\ \bibinfo {author} {\bibfnamefont {B.~L.}\ \bibnamefont {Weiss}},\
  }\href {http://dx.doi.org/10.1117/12.137584} {\bibfield  {journal} {\bibinfo
  {journal} {Proc. SPIE}\ }\textbf {\bibinfo {volume} {1675}},\ \bibinfo
  {pages} {98} (\bibinfo {year} {1992})}\BibitemShut {NoStop}%
\bibitem [{\citenamefont {Chuang}(2009)}]{Chuang}%
  \BibitemOpen
  \bibfield  {author} {\bibinfo {author} {\bibfnamefont {S.~L.}\ \bibnamefont
  {Chuang}},\ }\href@noop {} {\emph {\bibinfo {title} {{Physics of Photonic
  Devices}}}}\ (\bibinfo  {publisher} {Wiley},\ \bibinfo {year}
  {2009})\BibitemShut {NoStop}%
\bibitem [{\citenamefont {Piprek}(2003)}]{piprek-book03}%
  \BibitemOpen
  \bibfield  {author} {\bibinfo {author} {\bibfnamefont {J.}~\bibnamefont
  {Piprek}},\ }\href {\doibase 10.1016/B978-0-08-046978-2.50030-2} {\emph
  {\bibinfo {title} {{Semiconductor Optoelectronic Device (Ch 5)}}}}\ (\bibinfo
   {publisher} {Academic Press},\ \bibinfo {address} {Boston},\ \bibinfo {year}
  {2003})\ pp.\ \bibinfo {pages} {121 -- 139}\BibitemShut {NoStop}%
\bibitem [{\citenamefont {Sugawara}(1999)}]{sugawara-book99}%
  \BibitemOpen
  \bibfield  {author} {\bibinfo {author} {\bibfnamefont {M.}~\bibnamefont
  {Sugawara}},\ }in\ \href {\doibase
  http://dx.doi.org/10.1016/S0080-8784(08)62527-2} {\emph {\bibinfo {booktitle}
  {Self-Assembled InGaAs/GaAs Quantum Dots}}},\ \bibinfo {series}
  {Semiconductors and Semimetals}, Vol.~\bibinfo {volume} {60},\ \bibinfo
  {editor} {edited by\ \bibinfo {editor} {\bibfnamefont {R.}~\bibnamefont
  {Willardson}}\ and\ \bibinfo {editor} {\bibfnamefont {E.~R.}\ \bibnamefont
  {Weber}}}\ (\bibinfo  {publisher} {Elsevier},\ \bibinfo {year} {1999})\ pp.\
  \bibinfo {pages} {1 -- 116}\BibitemShut {NoStop}%
\bibitem [{\citenamefont {Fan}\ \emph {et~al.}(1996)\citenamefont {Fan},
  \citenamefont {Li}, \citenamefont {Chong},\ and\ \citenamefont
  {Xia}}]{fan96}%
  \BibitemOpen
  \bibfield  {author} {\bibinfo {author} {\bibfnamefont {W.~J.}\ \bibnamefont
  {Fan}}, \bibinfo {author} {\bibfnamefont {M.~F.}\ \bibnamefont {Li}},
  \bibinfo {author} {\bibfnamefont {T.~C.}\ \bibnamefont {Chong}}, \ and\
  \bibinfo {author} {\bibfnamefont {J.~B.}\ \bibnamefont {Xia}},\ }\href
  {\doibase 10.1063/1.363217} {\bibfield  {journal} {\bibinfo  {journal} {J.
  Appl. Phys.}\ }\textbf {\bibinfo {volume} {80}},\ \bibinfo {pages} {3471}
  (\bibinfo {year} {1996})}\BibitemShut {NoStop}%
\bibitem [{\citenamefont {Varshni}(1967)}]{varshni67}%
  \BibitemOpen
  \bibfield  {author} {\bibinfo {author} {\bibfnamefont {Y.~P.}\ \bibnamefont
  {Varshni}},\ }\href {\doibase http://dx.doi.org/10.1016/0031-8914(67)90062-6}
  {\bibfield  {journal} {\bibinfo  {journal} {Physica}\ }\textbf {\bibinfo
  {volume} {34}},\ \bibinfo {pages} {149 } (\bibinfo {year}
  {1967})}\BibitemShut {NoStop}%
\bibitem [{\citenamefont {Calderon}(2002)}]{calderon}%
  \BibitemOpen
  \bibfield  {author} {\bibinfo {author} {\bibfnamefont {I.~H.}\ \bibnamefont
  {Calderon}},\ }in\ \href@noop {} {\emph {\bibinfo {booktitle} {{Vol 12:
  II-VI} {S}emiconductor {M}aterials and their {A}pplications}}},\ \bibinfo
  {editor} {edited by\ \bibinfo {editor} {\bibfnamefont {M.~C.}\ \bibnamefont
  {Tamargo}}}\ (\bibinfo  {publisher} {Taylor and Francis},\ \bibinfo {address}
  {New York},\ \bibinfo {year} {2002})\BibitemShut {NoStop}%
\bibitem [{\citenamefont {Logothetidis}\ \emph {et~al.}(1986)\citenamefont
  {Logothetidis}, \citenamefont {Cardona}, \citenamefont {Lautenschlager},\
  and\ \citenamefont {Garriga}}]{logothetidis86}%
  \BibitemOpen
  \bibfield  {author} {\bibinfo {author} {\bibfnamefont {S.}~\bibnamefont
  {Logothetidis}}, \bibinfo {author} {\bibfnamefont {M.}~\bibnamefont
  {Cardona}}, \bibinfo {author} {\bibfnamefont {P.}~\bibnamefont
  {Lautenschlager}}, \ and\ \bibinfo {author} {\bibfnamefont {M.}~\bibnamefont
  {Garriga}},\ }\href {http://link.aps.org/doi/10.1103/PhysRevB.34.2458}
  {\bibfield  {journal} {\bibinfo  {journal} {Phys. Rev. B}\ }\textbf {\bibinfo
  {volume} {34}},\ \bibinfo {pages} {2458} (\bibinfo {year}
  {1986})}\BibitemShut {NoStop}%
\bibitem [{\citenamefont {Burov}\ \emph {et~al.}(2007)\citenamefont {Burov},
  \citenamefont {Lebedok}, \citenamefont {Kononenko}, \citenamefont
  {Ryabtsev},\ and\ \citenamefont {Ryabtsev}}]{burov07}%
  \BibitemOpen
  \bibfield  {author} {\bibinfo {author} {\bibfnamefont {L.~I.}\ \bibnamefont
  {Burov}}, \bibinfo {author} {\bibfnamefont {E.~V.}\ \bibnamefont {Lebedok}},
  \bibinfo {author} {\bibfnamefont {V.~K.}\ \bibnamefont {Kononenko}}, \bibinfo
  {author} {\bibfnamefont {A.~G.}\ \bibnamefont {Ryabtsev}}, \ and\ \bibinfo
  {author} {\bibfnamefont {G.~I.}\ \bibnamefont {Ryabtsev}},\ }\href {\doibase
  10.1007/s10812-007-0136-2} {\bibfield  {journal} {\bibinfo  {journal} {J.
  Appl. Spectroscopy}\ }\textbf {\bibinfo {volume} {74}},\ \bibinfo {pages}
  {878} (\bibinfo {year} {2007})}\BibitemShut {NoStop}%
\bibitem [{Note1()}]{Note1}%
  \BibitemOpen
  \bibinfo {note} {Later in Sec.\ \ref {subsec:expt-res} we will show that the
  experimental PL emission spectrum of 4 ML CdSe NPL at 30$^\circ $C (Fig.\
  \ref {fig:PL-Abs-TRPL}), has its peak emission at 512 nm corresponding to the
  transition energy of 2.42 eV predicted by our model.}\BibitemShut {Stop}%
\bibitem [{\citenamefont {Wei}\ and\ \citenamefont {Zunger}(1988)}]{wei88}%
  \BibitemOpen
  \bibfield  {author} {\bibinfo {author} {\bibfnamefont {S.-H.}\ \bibnamefont
  {Wei}}\ and\ \bibinfo {author} {\bibfnamefont {A.}~\bibnamefont {Zunger}},\
  }\href {\doibase 10.1103/PhysRevB.37.8958} {\bibfield  {journal} {\bibinfo
  {journal} {Phys. Rev. B}\ }\textbf {\bibinfo {volume} {37}},\ \bibinfo
  {pages} {8958} (\bibinfo {year} {1988})}\BibitemShut {NoStop}%
\bibitem [{\citenamefont {Olutas}\ \emph {et~al.}(2016)\citenamefont {Olutas},
  \citenamefont {Guzelturk}, \citenamefont {Kelestemur}, \citenamefont
  {Gungor},\ and\ \citenamefont {Demir}}]{olutas16}%
  \BibitemOpen
  \bibfield  {author} {\bibinfo {author} {\bibfnamefont {M.}~\bibnamefont
  {Olutas}}, \bibinfo {author} {\bibfnamefont {B.}~\bibnamefont {Guzelturk}},
  \bibinfo {author} {\bibfnamefont {Y.}~\bibnamefont {Kelestemur}}, \bibinfo
  {author} {\bibfnamefont {K.}~\bibnamefont {Gungor}}, \ and\ \bibinfo {author}
  {\bibfnamefont {H.~V.}\ \bibnamefont {Demir}},\ }\href {\doibase
  10.1002/adfm.201505108} {\bibfield  {journal} {\bibinfo  {journal} {Adv.
  Funct. Mater.}\ }\textbf {\bibinfo {volume} {26}},\ \bibinfo {pages} {2891}
  (\bibinfo {year} {2016})}\BibitemShut {NoStop}%
\bibitem [{\citenamefont {Bertrand}\ \emph {et~al.}(2016)\citenamefont
  {Bertrand}, \citenamefont {Polovitsyn}, \citenamefont {Christodoulou},
  \citenamefont {Khan},\ and\ \citenamefont {Moreels}}]{bertrand16}%
  \BibitemOpen
  \bibfield  {author} {\bibinfo {author} {\bibfnamefont {G.~H.~V.}\
  \bibnamefont {Bertrand}}, \bibinfo {author} {\bibfnamefont {A.}~\bibnamefont
  {Polovitsyn}}, \bibinfo {author} {\bibfnamefont {S.}~\bibnamefont
  {Christodoulou}}, \bibinfo {author} {\bibfnamefont {A.~H.}\ \bibnamefont
  {Khan}}, \ and\ \bibinfo {author} {\bibfnamefont {I.}~\bibnamefont
  {Moreels}},\ }\href {\doibase 10.1039/C6CC05705E} {\bibfield  {journal}
  {\bibinfo  {journal} {Chem. Commun.}\ }\textbf {\bibinfo {volume} {52}},\
  \bibinfo {pages} {11975} (\bibinfo {year} {2016})}\BibitemShut {NoStop}%
\bibitem [{\citenamefont {Achtstein}\ \emph {et~al.}(2015)\citenamefont
  {Achtstein}, \citenamefont {Antanovich}, \citenamefont {Prudnikau},
  \citenamefont {Scott}, \citenamefont {Woggon},\ and\ \citenamefont
  {Artemyev}}]{Achtstein15_jpcc}%
  \BibitemOpen
  \bibfield  {author} {\bibinfo {author} {\bibfnamefont {A.~W.}\ \bibnamefont
  {Achtstein}}, \bibinfo {author} {\bibfnamefont {A.}~\bibnamefont
  {Antanovich}}, \bibinfo {author} {\bibfnamefont {A.}~\bibnamefont
  {Prudnikau}}, \bibinfo {author} {\bibfnamefont {R.}~\bibnamefont {Scott}},
  \bibinfo {author} {\bibfnamefont {U.}~\bibnamefont {Woggon}}, \ and\ \bibinfo
  {author} {\bibfnamefont {M.}~\bibnamefont {Artemyev}},\ }\href {\doibase
  10.1021/acs.jpcc.5b06208} {\bibfield  {journal} {\bibinfo  {journal} {J.
  Phys. Chem. C}\ }\textbf {\bibinfo {volume} {119}},\ \bibinfo {pages} {20156}
  (\bibinfo {year} {2015})}\BibitemShut {NoStop}%
\bibitem [{\citenamefont {Tessier}\ \emph {et~al.}(2014)\citenamefont
  {Tessier}, \citenamefont {Spinicelli}, \citenamefont {Dupont}, \citenamefont
  {Patriarche}, \citenamefont {Ithurria},\ and\ \citenamefont
  {Dubertret}}]{tessier14}%
  \BibitemOpen
  \bibfield  {author} {\bibinfo {author} {\bibfnamefont {M.~D.}\ \bibnamefont
  {Tessier}}, \bibinfo {author} {\bibfnamefont {P.}~\bibnamefont {Spinicelli}},
  \bibinfo {author} {\bibfnamefont {D.}~\bibnamefont {Dupont}}, \bibinfo
  {author} {\bibfnamefont {G.}~\bibnamefont {Patriarche}}, \bibinfo {author}
  {\bibfnamefont {S.}~\bibnamefont {Ithurria}}, \ and\ \bibinfo {author}
  {\bibfnamefont {B.}~\bibnamefont {Dubertret}},\ }\href {\doibase
  10.1021/nl403746p} {\bibfield  {journal} {\bibinfo  {journal} {Nano Lett.}\
  }\textbf {\bibinfo {volume} {14}},\ \bibinfo {pages} {207} (\bibinfo {year}
  {2014})}\BibitemShut {NoStop}%
\bibitem [{\citenamefont {Mahler}\ \emph {et~al.}(2012)\citenamefont {Mahler},
  \citenamefont {Nadal}, \citenamefont {Bouet}, \citenamefont {Patriarche},\
  and\ \citenamefont {Dubertret}}]{Mahler12_jacs}%
  \BibitemOpen
  \bibfield  {author} {\bibinfo {author} {\bibfnamefont {B.}~\bibnamefont
  {Mahler}}, \bibinfo {author} {\bibfnamefont {B.}~\bibnamefont {Nadal}},
  \bibinfo {author} {\bibfnamefont {C.}~\bibnamefont {Bouet}}, \bibinfo
  {author} {\bibfnamefont {G.}~\bibnamefont {Patriarche}}, \ and\ \bibinfo
  {author} {\bibfnamefont {B.}~\bibnamefont {Dubertret}},\ }\href {\doibase
  10.1021/ja307944d} {\bibfield  {journal} {\bibinfo  {journal} {J. Am. Chem.
  Soc.}\ }\textbf {\bibinfo {volume} {134}},\ \bibinfo {pages} {18591}
  (\bibinfo {year} {2012})}\BibitemShut {NoStop}%
\bibitem [{\citenamefont {Kelestemur}\ \emph {et~al.}(2016)\citenamefont
  {Kelestemur}, \citenamefont {Guzelturk}, \citenamefont {Erdem}, \citenamefont
  {Olutas}, \citenamefont {Gungor},\ and\ \citenamefont
  {Demir}}]{Kelestemur_AFM16}%
  \BibitemOpen
  \bibfield  {author} {\bibinfo {author} {\bibfnamefont {Y.}~\bibnamefont
  {Kelestemur}}, \bibinfo {author} {\bibfnamefont {B.}~\bibnamefont
  {Guzelturk}}, \bibinfo {author} {\bibfnamefont {O.}~\bibnamefont {Erdem}},
  \bibinfo {author} {\bibfnamefont {M.}~\bibnamefont {Olutas}}, \bibinfo
  {author} {\bibfnamefont {K.}~\bibnamefont {Gungor}}, \ and\ \bibinfo {author}
  {\bibfnamefont {H.~V.}\ \bibnamefont {Demir}},\ }\href {\doibase
  10.1002/adfm.201600588} {\bibfield  {journal} {\bibinfo  {journal} {Adv.
  Funct. Mater.}\ }\textbf {\bibinfo {volume} {26}},\ \bibinfo {pages} {3570}
  (\bibinfo {year} {2016})}\BibitemShut {NoStop}%
\bibitem [{\citenamefont {Olutas}\ \emph {et~al.}(2015)\citenamefont {Olutas},
  \citenamefont {Guzelturk}, \citenamefont {Kelestemur}, \citenamefont
  {Yeltik}, \citenamefont {Delikanli},\ and\ \citenamefont {Demir}}]{olutas15}%
  \BibitemOpen
  \bibfield  {author} {\bibinfo {author} {\bibfnamefont {M.}~\bibnamefont
  {Olutas}}, \bibinfo {author} {\bibfnamefont {B.}~\bibnamefont {Guzelturk}},
  \bibinfo {author} {\bibfnamefont {Y.}~\bibnamefont {Kelestemur}}, \bibinfo
  {author} {\bibfnamefont {A.}~\bibnamefont {Yeltik}}, \bibinfo {author}
  {\bibfnamefont {S.}~\bibnamefont {Delikanli}}, \ and\ \bibinfo {author}
  {\bibfnamefont {H.~V.}\ \bibnamefont {Demir}},\ }\href {\doibase
  10.1021/acsnano.5b01927} {\bibfield  {journal} {\bibinfo  {journal} {ACS
  Nano}\ }\textbf {\bibinfo {volume} {9}},\ \bibinfo {pages} {5041} (\bibinfo
  {year} {2015})}\BibitemShut {NoStop}%
\bibitem [{\citenamefont {Guzelturk}\ \emph
  {et~al.}(2014{\natexlab{b}})\citenamefont {Guzelturk}, \citenamefont {Erdem},
  \citenamefont {Olutas}, \citenamefont {Kelestemur},\ and\ \citenamefont
  {Demir}}]{guzelturk_acsNano14}%
  \BibitemOpen
  \bibfield  {author} {\bibinfo {author} {\bibfnamefont {B.}~\bibnamefont
  {Guzelturk}}, \bibinfo {author} {\bibfnamefont {O.}~\bibnamefont {Erdem}},
  \bibinfo {author} {\bibfnamefont {M.}~\bibnamefont {Olutas}}, \bibinfo
  {author} {\bibfnamefont {Y.}~\bibnamefont {Kelestemur}}, \ and\ \bibinfo
  {author} {\bibfnamefont {H.~V.}\ \bibnamefont {Demir}},\ }\href {\doibase
  10.1021/nn5053734} {\bibfield  {journal} {\bibinfo  {journal} {ACS Nano}\
  }\textbf {\bibinfo {volume} {8}},\ \bibinfo {pages} {12524} (\bibinfo {year}
  {2014}{\natexlab{b}})}\BibitemShut {NoStop}%
\bibitem [{\citenamefont {Hutter}\ \emph {et~al.}(2014)\citenamefont {Hutter},
  \citenamefont {Bladt}, \citenamefont {Goris}, \citenamefont {Pietra},
  \citenamefont {van~der Bok}, \citenamefont {Boneschanscher}, \citenamefont
  {de~Mello~Donegá}, \citenamefont {Bals},\ and\ \citenamefont
  {Vanmaekelbergh}}]{Hutter14}%
  \BibitemOpen
  \bibfield  {author} {\bibinfo {author} {\bibfnamefont {E.~M.}\ \bibnamefont
  {Hutter}}, \bibinfo {author} {\bibfnamefont {E.}~\bibnamefont {Bladt}},
  \bibinfo {author} {\bibfnamefont {B.}~\bibnamefont {Goris}}, \bibinfo
  {author} {\bibfnamefont {F.}~\bibnamefont {Pietra}}, \bibinfo {author}
  {\bibfnamefont {J.~C.}\ \bibnamefont {van~der Bok}}, \bibinfo {author}
  {\bibfnamefont {M.~P.}\ \bibnamefont {Boneschanscher}}, \bibinfo {author}
  {\bibfnamefont {C.}~\bibnamefont {de~Mello~Donegá}}, \bibinfo {author}
  {\bibfnamefont {S.}~\bibnamefont {Bals}}, \ and\ \bibinfo {author}
  {\bibfnamefont {D.}~\bibnamefont {Vanmaekelbergh}},\ }\href {\doibase
  10.1021/nl5025744} {\bibfield  {journal} {\bibinfo  {journal} {Nano Lett.}\
  }\textbf {\bibinfo {volume} {14}},\ \bibinfo {pages} {6257} (\bibinfo {year}
  {2014})},\ \Eprint {http://arxiv.org/abs/http://dx.doi.org/10.1021/nl5025744}
  {http://dx.doi.org/10.1021/nl5025744} \BibitemShut {NoStop}%
\bibitem [{\citenamefont {Lakowicz}(2006)}]{Lakowicz-book06}%
  \BibitemOpen
  \bibfield  {author} {\bibinfo {author} {\bibfnamefont {J.~R.}\ \bibnamefont
  {Lakowicz}},\ }\href {\doibase 10.1007/978-0-387-46312-4} {\emph {\bibinfo
  {title} {{Principles of Fluorescence Spectroscopy}}}}\ (\bibinfo  {publisher}
  {Springer US},\ \bibinfo {address} {Boston},\ \bibinfo {year}
  {2006})\BibitemShut {NoStop}%
\bibitem [{\citenamefont {Erdem}\ \emph {et~al.}(2016)\citenamefont {Erdem},
  \citenamefont {Olutas}, \citenamefont {Guzelturk}, \citenamefont
  {Kelestemur},\ and\ \citenamefont {Demir}}]{Erdem16}%
  \BibitemOpen
  \bibfield  {author} {\bibinfo {author} {\bibfnamefont {O.}~\bibnamefont
  {Erdem}}, \bibinfo {author} {\bibfnamefont {M.}~\bibnamefont {Olutas}},
  \bibinfo {author} {\bibfnamefont {B.}~\bibnamefont {Guzelturk}}, \bibinfo
  {author} {\bibfnamefont {Y.}~\bibnamefont {Kelestemur}}, \ and\ \bibinfo
  {author} {\bibfnamefont {H.~V.}\ \bibnamefont {Demir}},\ }\href {\doibase
  10.1021/acs.jpclett.5b02763} {\bibfield  {journal} {\bibinfo  {journal} {J.
  Phys. Chem. Lett.}\ }\textbf {\bibinfo {volume} {7}},\ \bibinfo {pages} {548}
  (\bibinfo {year} {2016})}\BibitemShut {NoStop}%
\bibitem [{\citenamefont {Halder}\ \emph {et~al.}(2014)\citenamefont {Halder},
  \citenamefont {Pradhani}, \citenamefont {Sahoo}, \citenamefont {Satpati},\
  and\ \citenamefont {Rath}}]{halder14}%
  \BibitemOpen
  \bibfield  {author} {\bibinfo {author} {\bibfnamefont {O.}~\bibnamefont
  {Halder}}, \bibinfo {author} {\bibfnamefont {A.}~\bibnamefont {Pradhani}},
  \bibinfo {author} {\bibfnamefont {P.~K.}\ \bibnamefont {Sahoo}}, \bibinfo
  {author} {\bibfnamefont {B.}~\bibnamefont {Satpati}}, \ and\ \bibinfo
  {author} {\bibfnamefont {S.}~\bibnamefont {Rath}},\ }\href {\doibase
  10.1063/1.4875912} {\bibfield  {journal} {\bibinfo  {journal} {Appl. Phys.
  Lett.}\ }\textbf {\bibinfo {volume} {104}},\ \bibinfo {pages} {182109}
  (\bibinfo {year} {2014})}\BibitemShut {NoStop}%
\bibitem [{\citenamefont {Chia}\ \emph {et~al.}(2008)\citenamefont {Chia},
  \citenamefont {Yuan}, \citenamefont {Ku}, \citenamefont {Yang}, \citenamefont
  {Chou}, \citenamefont {Juang}, \citenamefont {Hsieh}, \citenamefont {Chiu},
  \citenamefont {Hsu},\ and\ \citenamefont {Jeng}}]{chia08}%
  \BibitemOpen
  \bibfield  {author} {\bibinfo {author} {\bibfnamefont {C.}~\bibnamefont
  {Chia}}, \bibinfo {author} {\bibfnamefont {C.}~\bibnamefont {Yuan}}, \bibinfo
  {author} {\bibfnamefont {J.}~\bibnamefont {Ku}}, \bibinfo {author}
  {\bibfnamefont {S.}~\bibnamefont {Yang}}, \bibinfo {author} {\bibfnamefont
  {W.}~\bibnamefont {Chou}}, \bibinfo {author} {\bibfnamefont {J.}~\bibnamefont
  {Juang}}, \bibinfo {author} {\bibfnamefont {S.}~\bibnamefont {Hsieh}},
  \bibinfo {author} {\bibfnamefont {K.}~\bibnamefont {Chiu}}, \bibinfo {author}
  {\bibfnamefont {J.}~\bibnamefont {Hsu}}, \ and\ \bibinfo {author}
  {\bibfnamefont {S.}~\bibnamefont {Jeng}},\ }\href
  {http://dx.doi.org/10.1016/j.jlumin.2007.06.003} {\bibfield  {journal}
  {\bibinfo  {journal} {J. Lumin.}\ }\textbf {\bibinfo {volume} {128}},\
  \bibinfo {pages} {123 } (\bibinfo {year} {2008})}\BibitemShut {NoStop}%
\bibitem [{\citenamefont {Rudin}, \citenamefont {Reinecke},\ and\ \citenamefont
  {Segall}(1990)}]{rudin90}%
  \BibitemOpen
  \bibfield  {author} {\bibinfo {author} {\bibfnamefont {S.}~\bibnamefont
  {Rudin}}, \bibinfo {author} {\bibfnamefont {T.~L.}\ \bibnamefont {Reinecke}},
  \ and\ \bibinfo {author} {\bibfnamefont {B.}~\bibnamefont {Segall}},\ }\href
  {\doibase 10.1103/PhysRevB.42.11218} {\bibfield  {journal} {\bibinfo
  {journal} {Phys. Rev. B}\ }\textbf {\bibinfo {volume} {42}},\ \bibinfo
  {pages} {11218} (\bibinfo {year} {1990})}\BibitemShut {NoStop}%
\bibitem [{\citenamefont {Perna}, \citenamefont {Capozzi},\ and\ \citenamefont
  {Ambrico}(1998)}]{perna98}%
  \BibitemOpen
  \bibfield  {author} {\bibinfo {author} {\bibfnamefont {G.}~\bibnamefont
  {Perna}}, \bibinfo {author} {\bibfnamefont {V.}~\bibnamefont {Capozzi}}, \
  and\ \bibinfo {author} {\bibfnamefont {M.}~\bibnamefont {Ambrico}},\ }\href
  {http://dx.doi.org/10.1063/1.367102} {\bibfield  {journal} {\bibinfo
  {journal} {J. Appl. Phys.}\ }\textbf {\bibinfo {volume} {83}},\ \bibinfo
  {pages} {3337} (\bibinfo {year} {1998})}\BibitemShut {NoStop}%
\bibitem [{\citenamefont {Takagahara}(1993)}]{takagahara93}%
  \BibitemOpen
  \bibfield  {author} {\bibinfo {author} {\bibfnamefont {T.}~\bibnamefont
  {Takagahara}},\ }\href {http://link.aps.org/doi/10.1103/PhysRevLett.71.3577}
  {\bibfield  {journal} {\bibinfo  {journal} {Phys. Rev. Lett.}\ }\textbf
  {\bibinfo {volume} {71}},\ \bibinfo {pages} {3577} (\bibinfo {year}
  {1993})}\BibitemShut {NoStop}%
\bibitem [{\citenamefont {Valerini}\ \emph {et~al.}(2005)\citenamefont
  {Valerini}, \citenamefont {Cret\'{\i}}, \citenamefont {Lomascolo},
  \citenamefont {Manna}, \citenamefont {Cingolani},\ and\ \citenamefont
  {Anni}}]{valerini05}%
  \BibitemOpen
  \bibfield  {author} {\bibinfo {author} {\bibfnamefont {D.}~\bibnamefont
  {Valerini}}, \bibinfo {author} {\bibfnamefont {A.}~\bibnamefont
  {Cret\'{\i}}}, \bibinfo {author} {\bibfnamefont {M.}~\bibnamefont
  {Lomascolo}}, \bibinfo {author} {\bibfnamefont {L.}~\bibnamefont {Manna}},
  \bibinfo {author} {\bibfnamefont {R.}~\bibnamefont {Cingolani}}, \ and\
  \bibinfo {author} {\bibfnamefont {M.}~\bibnamefont {Anni}},\ }\href
  {http://link.aps.org/doi/10.1103/PhysRevB.71.235409} {\bibfield  {journal}
  {\bibinfo  {journal} {Phys. Rev. B}\ }\textbf {\bibinfo {volume} {71}},\
  \bibinfo {pages} {235409} (\bibinfo {year} {2005})}\BibitemShut {NoStop}%
\bibitem [{\citenamefont {Leroux}\ \emph {et~al.}(1999)\citenamefont {Leroux},
  \citenamefont {Grandjean}, \citenamefont {Beaumont}, \citenamefont {Nataf},
  \citenamefont {Semond}, \citenamefont {Massies},\ and\ \citenamefont
  {Gibart}}]{leroux99}%
  \BibitemOpen
  \bibfield  {author} {\bibinfo {author} {\bibfnamefont {M.}~\bibnamefont
  {Leroux}}, \bibinfo {author} {\bibfnamefont {N.}~\bibnamefont {Grandjean}},
  \bibinfo {author} {\bibfnamefont {B.}~\bibnamefont {Beaumont}}, \bibinfo
  {author} {\bibfnamefont {G.}~\bibnamefont {Nataf}}, \bibinfo {author}
  {\bibfnamefont {F.}~\bibnamefont {Semond}}, \bibinfo {author} {\bibfnamefont
  {J.}~\bibnamefont {Massies}}, \ and\ \bibinfo {author} {\bibfnamefont
  {P.}~\bibnamefont {Gibart}},\ }\href {http://dx.doi.org/10.1063/1.371242}
  {\bibfield  {journal} {\bibinfo  {journal} {J. Appl. Phys.}\ }\textbf
  {\bibinfo {volume} {86}},\ \bibinfo {pages} {3721} (\bibinfo {year}
  {1999})}\BibitemShut {NoStop}%
\bibitem [{\citenamefont {Cai}\ \emph {et~al.}(2013)\citenamefont {Cai},
  \citenamefont {Martin}, \citenamefont {Shea-Rohwer}, \citenamefont {Gong},\
  and\ \citenamefont {Kelley}}]{cai13}%
  \BibitemOpen
  \bibfield  {author} {\bibinfo {author} {\bibfnamefont {X.}~\bibnamefont
  {Cai}}, \bibinfo {author} {\bibfnamefont {J.~E.}\ \bibnamefont {Martin}},
  \bibinfo {author} {\bibfnamefont {L.~E.}\ \bibnamefont {Shea-Rohwer}},
  \bibinfo {author} {\bibfnamefont {K.}~\bibnamefont {Gong}}, \ and\ \bibinfo
  {author} {\bibfnamefont {D.~F.}\ \bibnamefont {Kelley}},\ }\href
  {http://dx.doi.org/10.1021/jp400688g} {\bibfield  {journal} {\bibinfo
  {journal} {J. Phys. Chem. C}\ }\textbf {\bibinfo {volume} {117}},\ \bibinfo
  {pages} {7902} (\bibinfo {year} {2013})}\BibitemShut {NoStop}%
\bibitem [{\citenamefont {Liu}\ \emph {et~al.}(2006)\citenamefont {Liu},
  \citenamefont {Huang}, \citenamefont {Wang}, \citenamefont {Wang},
  \citenamefont {Li}, \citenamefont {Zhao},\ and\ \citenamefont {Luo}}]{liu06}%
  \BibitemOpen
  \bibfield  {author} {\bibinfo {author} {\bibfnamefont {T.-C.}\ \bibnamefont
  {Liu}}, \bibinfo {author} {\bibfnamefont {Z.-L.}\ \bibnamefont {Huang}},
  \bibinfo {author} {\bibfnamefont {H.-Q.}\ \bibnamefont {Wang}}, \bibinfo
  {author} {\bibfnamefont {J.-H.}\ \bibnamefont {Wang}}, \bibinfo {author}
  {\bibfnamefont {X.-Q.}\ \bibnamefont {Li}}, \bibinfo {author} {\bibfnamefont
  {Y.-D.}\ \bibnamefont {Zhao}}, \ and\ \bibinfo {author} {\bibfnamefont
  {Q.-M.}\ \bibnamefont {Luo}},\ }\href {\doibase 10.1016/j.aca.2005.11.053}
  {\bibfield  {journal} {\bibinfo  {journal} {Anal. Chim. Acta}\ }\textbf
  {\bibinfo {volume} {559}},\ \bibinfo {pages} {120 } (\bibinfo {year}
  {2006})}\BibitemShut {NoStop}%
\end{thebibliography}%

\end{document}